\pdfoutput=1

\newcommand*{\ATLASLATEXPATH}{latex/}

\documentclass[cernpreprint, texlive=2016, UKenglish]{\ATLASLATEXPATH atlasdoc}

\usepackage[full,subcaption,backend=biber]{\ATLASLATEXPATH atlaspackage}
\usepackage{\ATLASLATEXPATH atlasbiblatex}

\usepackage[hepprocess=true]{\ATLASLATEXPATH atlasphysics}

\graphicspath{{logos/}}

\usepackage{feynmp-auto}
\usepackage{multirow}
\usepackage{lscape}
\usepackage{longtable}
\usepackage{float}
\usepackage[normalem]{ulem}
\usepackage{cleveref}

\usepackage{EXOT-2017-22-PAPER-defs}

\AtlasTitle{Search for excited electrons singly produced in proton--proton collisions at \rts~=~13~\TeV\ with the ATLAS experiment at the LHC}

\AtlasAbstract{
A search for excited electrons produced in $pp$ collisions at \rts~=~13~\TeV\ via a contact interaction \qqbartoeestar is presented. The search uses 36.1~\ifb\ of data collected in 2015 and 2016 by the ATLAS experiment at the Large Hadron Collider. Decays of the excited electron into an electron and a pair of quarks (\eqqbar) are targeted in final states with two electrons and two hadronic jets, and decays via a gauge interaction into a neutrino and a \Wboson boson (\nuW) are probed in final states with an electron, missing transverse momentum, and a large-radius jet consistent with a hadronically decaying \Wboson boson. No significant excess is observed over the expected backgrounds. Upper limits are calculated for the $\pp\to\eestartoeeqqbar$ and $\pp\to\eestartoenuW$ production cross sections as a function of the excited electron mass \mestar at 95\% confidence level. The limits are translated into lower bounds on the compositeness scale parameter \Lamb of the model as a function of \mestar. For $\mestar<0.5$~\TeV, the lower bound for \Lamb is 11~\TeV.  In the special case of $\mestar=\Lamb$, the values of $\mestar<4.8$~\TeV\ are excluded. The presented limits on \Lamb are more stringent than those obtained in previous searches.
}

\author{The ATLAS Collaboration}

\AtlasRefCode{EXOT-2017-22}

\PreprintIdNumber{CERN-EP-2019-021}

\AtlasJournal{Eur. Phys. J. C}

\AtlasJournalRef{Eur. Phys. J. C 79 (2019) 803}

\AtlasDOI{10.1140/epjc/s10052-019-7295-1}

\AtlasCoverSupportingNote{Supporting internal document}{https://cds.cern.ch/record/2152378}

\hypersetup{pdftitle={ATLAS document},pdfauthor={The ATLAS Collaboration}}

\begin{document}

\maketitle

\section{Introduction}\label{sec:Introduction}

Excited leptons appear in a number of composite models~\cite{Pati:1974yy, Kayser:1981nw, Eichten:1983hw, Cabibbo:1983bk, Hagiwara:1985wt, Baur:1989kv} seeking to explain the existence of the three generations of quarks and leptons in the Standard Model (SM). This analysis uses the model presented in Ref.~\cite{Baur:1989kv} as a benchmark. 
The composite models introduce new constituent particles called preons that bind at a high scale \Lamb to form SM fermions and their excited states.
The preon bound states are mapped into representations of the $SU(2)\times U(1)$ SM gauge group. The SM fermions are identified as a set of left- and right-handed chiral states protected by the $SU(2)$ symmetry from obtaining masses of the order of \Lamb~\cite{Baur:1989kv}.
The remaining vector-like states, $SU(2)$ doublets and singlets, acquire masses of the order of \Lamb and are thus interpreted as excited fermions.
The effective Lagrangian introduces four-fermion contact-interaction (CI) terms (Eqs.~(\ref{eq:lcontact}) and (\ref{eq:jmu})) and gauge-mediated (GM) currents (Eq.~(\ref{eq:lgauge})):

\begin{equation}
\Delta\mathcal{L}_\mathrm{CI}=\frac{2\pi}{\Lambda^2}j^{\mu}j_{\mu}
\label{eq:lcontact}
\end{equation}

\vspace*{-2ex}
\begin{equation}
j_{\mu}=\bar{f_\mathrm{L}}\gamma_{\mu}f_\mathrm{L}+\bar{f^*_\mathrm{L}}\gamma_{\mu}f^*_\mathrm{L}+ \left( \bar{f^*_\mathrm{L}}\gamma_{\mu}f_\mathrm{L}+\mathrm{H.C.} \right)
\label{eq:jmu}
\end{equation}

\vspace*{-2ex}
\begin{equation}
\Delta\mathcal{L}_\mathrm{GM}=\frac{1}{2\Lambda}\bar{f^*_\mathrm{R}}\sigma^{\mu\nu}\left[g\frac{\tau}{2}W_{\mu\nu}+g'\frac{Y}{2}B_{\mu\nu}\right]f_\mathrm{L} + \mathrm{H.C.}
\label{eq:lgauge}
\end{equation}

\begin{figure}[htbp]
	\centering
	\includegraphics[width=0.48\textwidth]{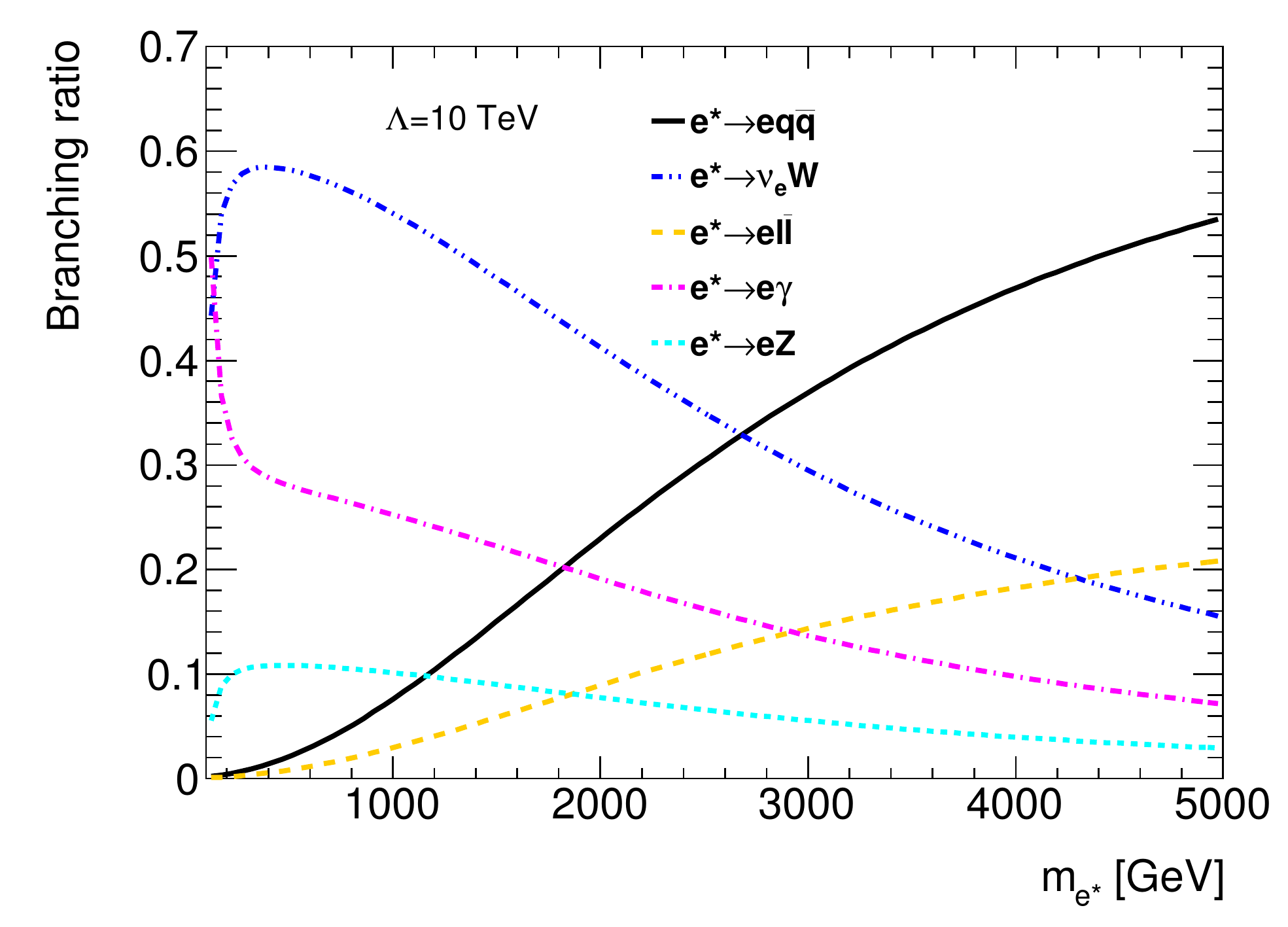}
	\caption{Branching ratios for excited electrons as a function of \mestar. The scale $\Lambda$ is set to 10~\TeV.}
	\label{fig:BRs}
\end{figure}

Here, $f = \ell, q$ and $f^* = \ell^*, q^*$ denote SM and excited leptons and quarks, and the subscripts L and R stand for left- and right-handed components of the fermion field $f$, respectively. The $j_{\mu}$ term is the fermion current of $f$ and $f^*$. The $W_{\mu\nu}$ and $B_{\mu\nu}$ are the field-strength tensors of the $SU(2)$ and $U(1)$ gauge fields, and $g$ and $g'$ are the corresponding coupling constants of the electroweak theory.
The left- and right-handed excited fermions are both $SU(2)$ doublets, with the weak hypercharge $Y$ such that $f^*$ electric charges coincide with the ones of their ground states $f$. The weak hypercharge $Y$ of the $\lstar_\mathrm{L,R}$ doublet is $-1$, so that its isospin $T_3 = -1/2$ component represents an excited lepton with electric charge $Q = -1$. Therefore, the excited lepton model introduces two unknown parameters relevant for this analysis, the excited lepton mass \mestar and the compositeness scale \Lamb, which define 
the preferred search channels and kinematic properties of the final states.
The four-fermion CI terms are suppressed by $1/\Lamb^2$ implying the parton-level $e^*$ production cross section growing proportionally to $\hat{s}$.
The considered models allow only left-handed currents in the contact-interaction terms, and all dimensionless couplings defining the relative strength of the residual interactions are set to unity~\cite{Baur:1989kv}. The restriction $\mestar<\Lamb$ follows from unitarity constraints on the contact interactions~\cite{Baur:1989kv, Anders:1993dk}. 
Branching ratios ($\mathcal{B}$) for excited electrons as functions of \mestar for the case of $\Lambda=10$~\TeV are presented in \Cref{fig:BRs}.
Gauge-mediated decays dominate at $\mestar \ll \Lambda$ while the decay via a contact interaction becomes dominant for $\mestar \gtrsim \Lambda / 3$.
\par

This article presents a search for excited electrons singly produced in \pp collisions at \rts~=~13~\TeV\ via a contact interaction \qqbartoeestar and decaying either to an electron and a pair of quarks (\eqqbar) via a contact interaction or to a neutrino and a \Wboson boson (\nuW) via a gauge interaction, depicted in \Cref{fig:FD:ELtolqq,fig:FD:ELtonuW}, respectively. 
Given the sensitivity of the search, the contribution of gauge-mediated production of the excited electrons is non-negligible relative to the contact-interaction production only for $\mestar < 200$~\GeV~\cite{Aad:2014hja} 
and thus neglected. 
The search uses 36.1~\ifb\ of data collected in 2015 and 2016 by the ATLAS experiment~\cite{PERF-2007-01} at the Large Hadron Collider (LHC).

\begin{figure}[htbp]
	\centering
	\begin{subfigure}[l]{.44\textwidth}
		\includegraphics[width=1\textwidth]{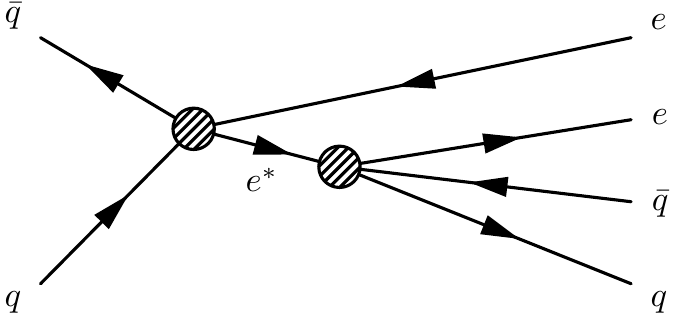}
		\caption{}
		\label{fig:FD:ELtolqq}
	\end{subfigure}
	\hfill
	\begin{subfigure}[r]{.44\textwidth}
		\includegraphics[width=1\textwidth]{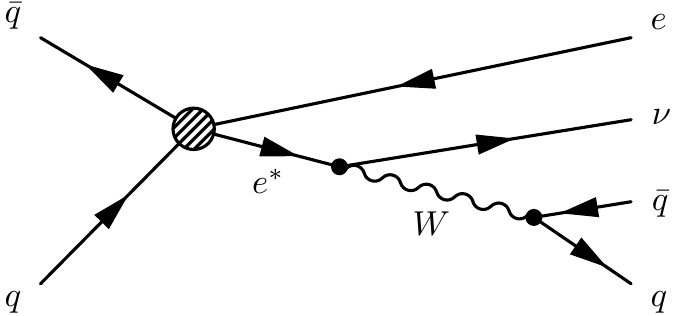}
		\caption{}
		\label{fig:FD:ELtonuW}
	\end{subfigure}
	\caption{Feynman diagrams for (a) \eestartoeeqqbar  and (b) \eestartoenuW.}
	\label{fig:FD}
\end{figure}

The present search uses two experimental channels. The first channel targets the production of excited electrons via a contact interaction \qqbartoeestar and their decay via a contact interaction \estartoeqqbar, resulting in two energetic electrons and at least two hadronic jets $j$. In the second channel, the excited electrons are produced via a contact interaction as well, but their decay is via a gauge-mediated interaction into a \Wboson and a $\nu$, where the \Wboson boson decays hadronically, yielding an \eestartoenuqqbar final state. Experimentally, this gives final states with exactly one energetic electron, a large-radius (large-$R$) jet $J$ produced by two collimated quarks, and missing transverse momentum. 
The large-$R$ jet approach is sufficient for the current analysis, as the analysis selection with two resolved jets has minor efficiency. 
In the following, the final states resulting from contact- and gauge-mediated decays of singly produced \estar are denoted by \eejj and \enuJ, respectively.
The combination of the two channels maximizes the sensitivity of the search for all $\mestar/\Lambda$ values.
For possible reinterpretations, the results are also presented in terms of model-independent upper limits on the number of signal events and on the visible signal cross section.
\par
Previous searches for excited leptons were carried out at LEP \cite{Buskulic:1996tw, Abbiendi:1999sa, Achard:2003hd, Abdallah:2004yn}, HERA \cite{Chekanov:2001xk, Aaron:2008cy}, the Tevatron \cite{Acosta:2004ri, Abulencia:2006hj, Abazov:2006wq, Abazov:2008ag}, and the LHC \cite{ATLAS:2012af, Aad:2013jja, Aad:2014hja, Aad:2016giq, Chatrchyan:2011pg, CMS:2012ad, Khachatryan:2015scf, Sirunyan:2018zzr}. No evidence of excited leptons was found and bounds were set on \mestar, which is limited to be greater than 3~\TeV\ for the compositeness scale $\Lamb = \mestar$~\cite{Aad:2013jja}.
\par

\section{ATLAS detector}\label{sec:ATLAS}

The ATLAS detector \cite{PERF-2007-01} is a multipurpose detector with a forward--backward symmetric cylindrical geometry and nearly 4$\pi$ coverage in solid  angle.\footnote{ATLAS uses a right-handed coordinate system with its origin at the nominal interaction point (IP) in the centre of the detector and the $z$-axis along the beam pipe. The $x$-axis points from the IP to the centre of the LHC ring, and the $y$-axis points upward. Cylindrical coordinates $(r,\phi)$ are used in the transverse plane, $\phi$ being the azimuthal angle around the $z$-axis. The pseudorapidity is defined in terms of the polar angle $\theta$ as $\eta=-\ln\tan(\theta/2)$. Angular distance is measured in units of $\Delta R \equiv \sqrt{(\Delta\eta)^{2} + (\Delta\phi)^{2}}$.} 
The three major subcomponents of ATLAS are the tracking detector, the calorimeter, and the muon spectrometer.  Charged-particle tracks and vertices are reconstructed by the inner detector (ID) tracking system, comprising silicon pixel (including the newly installed innermost pixel layer \cite{Capeans:2010jnh,Abbott:2018ikt}) and silicon microstrip detectors covering the pseudorapidity range $|\eta|$~$<$~2.5, and a straw-tube tracker that covers $|\eta|$~$<$~2.0.  The ID is immersed in a homogeneous 2\,T magnetic field provided by a solenoid. The energies of electrons, photons, and jets are measured with sampling calorimeters. The ATLAS calorimeter system covers a pseudorapidity range of $|\eta|$~$<$~4.9.  Within the region $|\eta|$~$<$~3.2, electromagnetic (EM) calorimetry is performed with barrel and endcap high-granularity lead/liquid argon (LAr) calorimeters, with an additional thin LAr presampler covering $|\eta|$~$<$~1.8 to correct for energy loss in material upstream of the calorimeters. Hadronic calorimetry is performed with a steel/scintillator-tile calorimeter, segmented into three barrel structures within $|\eta|$~$<$~1.7, and two copper/LAr endcap calorimeters. The forward region (3.1~$<$~$|\eta|$~$<$~4.9) is instrumented with a LAr calorimeter with copper and tungsten absorbers for EM and hadronic energy measurements, respectively. Surrounding the calorimeters is a muon spectrometer (MS) with superconducting air-core toroidal magnets. 
The field integral of the toroids ranges between \num{2.0} and \SI{6.0}{\tesla\metre} across most of the detector. 
The MS includes three stations of precision tracking chambers covering $|\eta|$~$<$~2.7 to measure the curvature of tracks. The MS also contains detectors with triggering capabilities covering $|\eta|$~$<$~2.4 to provide fast muon identification and momentum measurements.\par

The ATLAS two-level trigger system selects events as described in Ref.~\cite{Aaboud:2016leb}.
The first-level trigger is hardware-based while the second, high-level trigger is implemented in software and employs algorithms similar to those used offline in the full event reconstruction.

\section{Data and simulated event samples}\label{sec:DataAndMC}

The analysis uses the $pp$ collision data recorded by the ATLAS detector in 2015 and 2016 at \rts~=~13~\TeV\ with a 25~ns bunch spacing. The total integrated luminosity collected in the data-taking periods with normal operation of the relevant detector subsystems is 36.1~\ifb. To further improve the data quality, events containing noise bursts or coherent noise in the calorimeters, as well as incompletely recorded events, are excluded.
\par
Events for the \eejj channel were recorded using di-electron triggers with transverse energy \ET thresholds of 12 and 17~\GeV\ for both electrons in 2015 and 2016, respectively. For the \enuJ channel, events must pass at least one of the two single-electron trigger requirements with thresholds set at \ET~=~60 or 120~\GeV\ in 2015, and \ET~=~60 or 140~\GeV\ in 2016. 
Combining the lower-threshold trigger with the one with a higher threshold but looser identification requirements, results in single-electron trigger efficiencies typically exceeding 90\% for the electrons in the phase space considered in the analysis~\cite{Aaboud:2016leb}.
Events with an \emjj final state are used for background studies in the \eejj channel and are selected using a combination of the two single-muon triggers with the transverse momentum \pT thresholds of 26 and 50~\GeV.
\par
Selected events contain proton--proton collisions in the same or neighboring bunch crossing (pile-up). 
The events used in the analysis contain 24 pile-up interactions on average, resulting in multiple interaction vertices in an event.
The primary vertex (PV) is defined as the vertex with the highest $\Sigma p_\mathrm{T}^2$ of charged-particle tracks. This PV must have at least two tracks with the transverse momentum \pT>~400~\MeV.\par
\par
The signal samples were simulated by \PYTHIA8.210~\cite{Sjostrand:2014zea}, using a leading-order (LO) matrix element (ME), the \NNPDF23LO \cite{Ball:2012cx} set of parton distribution functions (PDFs) and the \A14~\cite{a14tunes} set of tuned parameters. The \estar widths for the simulated signal samples were derived from \CalcHEP 3.6.25~\cite{Belyaev:2012qa}, which takes into account phase-space effects due to quark masses. The samples were generated for a compositeness scale \Lamb~=~5~\TeV\ and masses of excited electrons ranging from 100~\GeV\ to 4~\TeV. 
The effect of a finite \Lamb-dependent \estar width on the analysis is negligible for $m_{\estar} < \Lamb$.
\par
As shown in \Cref{sec:BackgroundComposition}, the dominant backgrounds in the \eejj and \enuJ channels are from \zjets and \wjets production, respectively. The sub-leading background in both channels is from \ttbar production, followed by single-top and diboson production.
The estimation of background processes involving prompt leptons from $W$ and $Z/\gamma^*$ decays relies on simulated event samples.
\par
The \zjets and \wjets processes were simulated using \SHERPA~2.2.1~\cite{Gleisberg:2008ta}. Parton-level final states with up to two partons produced along with the \Zboson and \Wboson bosons were generated at next-to-leading order (NLO), and those with three or four partons were generated at LO, using the \textsc{OpenLoops}~\cite{Cascioli:2011va} and \textsc{Comix}~\cite{Gleisberg:2008fv} for the NLO and LO cases, respectively. Double counting of events with the same partonic final state generated by various combinations of the ME and parton shower (PS) was eliminated according to the ME+PS@NLO prescription~\cite{Hoeche:2012yf}. The NNPDF~3.0~\cite{Ball:2014uwa} set of PDFs was used.
The \zjets and \wjets simulated event samples were normalized to the next-to-next-to-leading-order (NNLO) inclusive cross sections computed with the FEWZ program~\cite{Li:2012wna}.
\par
The \ttbar simulated event samples were generated at NLO accuracy in the strong coupling constant using \POWHEGBOX~v2~\cite{Nason:2004rx,Frixione:2007vw,Frixione:2007nw,Alioli:2010xd}, with the top-quark spin correlations preserved, and the CT10~\cite{Lai:2010vv} PDF set. Electroweak $s$- and $t$-channel single-top-quark events as well as events with a single top-quark produced in association with a $W$ boson were generated using \POWHEGBOX~v1~\cite{Alioli:2009je,Re:2010bp}. 
Parton showering, hadronization and the underlying event were handled by \PYTHIA8.210 for \ttbar production and by \PYTHIA6.428~\cite{Sjostrand:2006za} for single-top production. \PYTHIA8.210 and \PYTHIA6.428 used the \A14 and Perugia~2012~\cite{Skands:2010ak} sets of tuned parameters, respectively. The \ttbar simulated event sample was normalized to the inclusive cross section calculated using the Top++~v2.0~\cite{Czakon:2011xx} at NNLO accuracy in the strong coupling constant, 
with soft gluon emission accounted for in the next-to-next-to-leading logarithmic order (NNLL).
The single-top simulated event samples were normalized to the cross sections computed at NLO+NNLL accuracy~\cite{Kidonakis:2011wy}.
\par
The $ZZ$, $ZW$ and $WW$ simulated event samples were generated using \SHERPA~2.2.1. Events containing zero or one final-state parton were generated using an NLO ME. Events with two or three recoiling quarks or gluons were generated with a LO ME. The NNPDF~3.0 PDF set was used. The event generator cross sections are used in this case.
\par 
Decays of $b$- and $c$-hadrons in the simulated event samples of \ttbar, single-top, and signal processes were handled by EvtGen~v1.2.0~\cite{Lange:2001uf}.
\par
The pile-up interactions are described by overlaying minimum-bias events on each simulated signal or background event. The minimum-bias events were generated with \PYTHIA~8.186~\cite{Sjostrand:2007gs} with the \textsc{A2}~\cite{ATL-PHYS-PUB-2012-003} set of tuned parameters and the \MSTW2008LO~\cite{Martin:2009iq} PDFs. The distribution of the average number of interactions per bunch crossing in simulated event samples is reweighted to match the observed data.
\par
All the simulated event samples were passed through a simulation of the ATLAS detector~\cite{Aad:2010ah}. The detector response was obtained from a detector model that uses \textsc{Geant}4~\cite{Agostinelli:2002hh}.
For the simulation of the \eestartoeeqqbar signal samples, \textsc{Geant}4 based inner detector simulation was combined with a parameterized calorimeter simulation~\cite{Aad:2010ah}. The simulated event samples were processed with the same reconstruction software as used for data.
\par

\section{Object and event selection}\label{sec:EventSelection}

Events satisfying basic quality, trigger and vertex requirements are selected for the analysis using the criteria applied to electrons, muons, hadronic jets, and their kinematic quantities. The looser \emph{baseline} selections are applied at stages which aim to eliminate double counting of detected objects (electrons, muons, jets, tracks, vertices, etc.) in an event and double-counting of events in the two analysis channels. The tighter \emph{final} selection defines objects used in the analysis. In the following, both the \emph{baseline} and \emph{final} object selections are specified in \Cref{tab:ObjDefChain}, and the order of the event criteria applied in the analysis is given in \Cref{tab:EvtSelChain}. These selections form the \emph{preselection} stage.
\par
An electron candidate is reconstructed as  a clustered energy deposition in the calorimeter matched to a track from the ID~\cite{Aaboud:2019ynx}.
The direction of an electron is taken from its track and the energy is measured from the EM cluster. The energy is corrected for losses in the material before the calorimeter and for leakage outside of the cluster~\cite{Aaboud:2018ugz}.
The coverage of the ID limits the pseudorapidity of electrons to $|\eta|<2.47$. Electrons with $1.37<|\eta|<1.52$ are excluded because they point to the barrel-to-endcap transition regions.
To reject electron candidates originating from hadronic jets and photon conversions, electrons are required to satisfy a set of likelihood-based identification criteria determined by variables characterizing longitudinal and lateral calorimeter shower shapes, ID track properties, and track--cluster matching. These criteria are referred to, in order of increasing background rejection, as \emph{loose}, \emph{medium} and \emph{tight} and are defined so that an electron satisfying a tighter criterion always satisfies looser ones. The \emph{loose} identification is approximately 95\% efficient for prompt electrons with \pT > 30~\GeV. In the same \pT range, signal efficiency for \emph{medium} identification is greater than 90\%. The efficiency of \emph{tight} identification is greater than 85\% for prompt electrons with \pT > 65~\GeV~\cite{Aaboud:2019ynx}.
Further rejection of background is achieved by applying EM calorimeter and ID isolation requirements~\cite{Aaboud:2019ynx}. 
The \emph{loose} isolation requirement applied in this analysis is designed to achieve $99$\% selection efficiency for prompt electrons.
Electrons originating from the primary interaction vertex are selected by requiring the reconstructed electron track to have a transverse impact parameter significance $|d_0|/\sigma_{d_0} < 5$, where $\sigma_{d_0}$ is the uncertainty in the transverse impact parameter, and a longitudinal impact parameter $|z_0\sin\theta| < 0.5$~mm.
\par
Muons are reconstructed using a combined fit of tracks measured with the ID and MS. Muons from in-flight decays of charged hadrons are suppressed with the \emph{medium} set of identification requirements~\cite{Aad:2016jkr}. The muon identification efficiency exceeds 96\% for prompt muons with \pT > 20~\GeV. Muons are also subject to a \emph{loose} isolation requirement that uses ID tracks~\cite{Aad:2016jkr} and is 99\% efficient for prompt muons at any relevant \pT and $\eta$. Muons are further required to originate from the primary vertex by imposing the same criteria as for electrons on the ID track's longitudinal impact parameter and transverse impact parameter significance less then 3.
\par
Hadronic jets are reconstructed from clustered energy deposits in the calorimeters using the anti-$k_t$ algorithm~\cite{Cacciari:2008gp} with radius parameters $R = 0.4$ and $R = 1.0$. 
The reconstructed jets with $R = 1.0$ are trimmed~\cite{Krohn2010:JetTrimming} to reduce contributions from pile-up interactions and underlying event by reclustering the jet constituents into subjets using a $k_t$ algorithm with $R = 0.2$ and removing subjets carrying less than 5\% of the boosted jet's \pT.
Jet calibrations are applied as described in Refs.~\cite{Aaboud:2017jcu,Aaboud:2018kfi}.
\par
An event is removed if it contains a jet reconstructed with $R=0.4$ and originating from non-collision backgrounds, which is identified either by a substantial fraction of the jet energy being deposited in known noisy calorimeter cells or by a low fraction of the jet energy being carried by charged particles originating from the primary vertex and lying within a $\Delta R=0.4$ cone around the jet axis~\cite{TheATLAScollaboration:2015ofz}. Rejection of pile-up jets with $|\eta| < 2.4$ and \pT< 60~\GeV\ is achieved using a jet-vertex-tagger (JVT) discriminant~\cite{Aad:2015ina} quantifying the relative probability for a jet to originate from the primary vertex.
\par
The $R=0.4$ jets containing $b$-hadrons ($b$-jets) are identified using the multivariate $b$-tagging algorithm \emph{MV2c10}~\cite{ATL-PHYS-PUB-2016-012} based on impact parameters of tracks within the jet cone and positions of secondary decay vertices~\cite{Aad:2015ydr}. 
The $b$-tagging efficiency is 77\% as measured in simulated \ttbar event samples~\cite{Aaboud:2018xwy}.
\par
To discriminate boosted jets originating from \Wboson boson decays from those produced through strong interactions, the jet mass obtained by combining measurements from the calorimeter and tracking systems and the substructure variable $D^{\beta=1}_2$~\cite{FatJetSubstructure,Larkoski:JetObservables} are used.  The function $D^{\beta=1}_2$ is a ratio of three- to two-point correlation functions based on the \pT values and pairwise $\Delta R$ separations of jet constituents. The $D^{\beta=1}_2$ variable is specifically sensitive to a two-prong substructure within a jet and tends to zero in a two-body decay limit. A boosted jet is tagged as a \Wboson candidate if its mass falls within a certain mass window around $m_W$ and its $D^{\beta=1}_2$ value is sufficiently low. For the \Wboson-tagging procedure the mass window and the upper bound placed on $D^{\beta=1}_2$ are tuned, depending on the jet \pT, to reach a nominal 50\% signal efficiency (\Wboson-tag50) with a multi-jet background rejection factor of 40--80~\cite{Aaboud:2018psm,ATL-PHYS-PUB-2015-033}. 
The jet energy and mass are both calibrated prior to applying the \Wboson-tagging discriminant. At the \emph{preselection} level, only the upper bound on $D^{\beta=1}_2$ corresponding to \Wboson-tag50 is imposed.
\par
The missing transverse momentum, with magnitude \etmiss, is calculated as the negative vector sum of all reconstructed objects associated with the primary vertex. 
This includes calibrated electrons, muons, and $R=0.4$ jets, and a track-based soft term (TST) using ID tracks not associated with the preselected hard objects~\cite{Aaboud:2018tkc}. The TST is built from tracks with \pT > 400~\MeV\ and $|\eta| < 2.5$ which have a sufficient number of hits in the ID, a good fit quality, and an origin consistent with the primary vertex.
\par
Double counting of electrons, muons, and jets reconstructed by more than one lepton and/or jet algorithm as well as misreconstruction of distinct physics objects produced in close proximity are resolved by the overlap removal procedure. The procedure is applied to the \emph{baseline} objects in the following order:
\begin{itemize}
	\item electron--electron: if two electrons share an ID track then the lower quality electron is removed; if both electrons are of the same quality then the lower-\pT electron is removed;
	\item electron--muon: remove the electron which shares an ID track with the muon;
	\item electron--jet with $R=0.4$: remove the jet if $\Delta{}R(e,\mathrm{jet}) < 0.2$ and, in the \enuJ channel only, the jet is not $b$-tagged;
	      after repeating this step for all pairs of electrons and surviving jets, electrons within $\Delta{}R = 0.4$ of a jet are removed;
	\item muon--jet with $R=0.4$: if $\Delta{}R(\mu,\mathrm{jet}) < 0.2$ and the jet has less than three ID tracks originating from the muon production vertex and, in the \enuJ channel only, the jet is not $b$-tagged, then the jet is removed; 
	after repeating this step for all pairs of muons and surviving jets, muons within $\Delta{}R = 0.4$ of a jet are removed.
\end{itemize}
The second overlap removal procedure applied only in the \enuJ channel involves \emph{baseline} electrons and \emph{final} boosted jets. The boosted jet is removed if a \emph{baseline} electron is present within $\Delta{}R=1.0$ of the boosted jet direction.
\par

\begin{table}[htbp]
\caption{Object definitions in the \eejj and \enuJ channels.
	 Muon selections are given in parentheses.
        }
\label{tab:ObjDefChain}
\begin{center}
\begin{tabular}{|c|c|c|c|}
\hline
Selection type & Objects & \eejj  & \enuJ \\
\hline
\multirow{10}{*}{\emph{Baseline}} 
	& \multirow{5}{*}{Electrons}
		& $\pT>30$~\GeV\ ($>40$~\GeV) & $\pT>40$~\GeV \\
	&	& \multicolumn{2}{c|}{$|\eta| < 2.47$, excluding $1.37<|\eta|<1.52$ ($< 2.5$)}\\
	&	& \multicolumn{2}{c|}{quality \emph{loose} (\emph{medium})}\\
	&	(muons) & \multicolumn{2}{c|}{no isolation (\emph{loose} isolation with ID tracks)}\\
	&	& \multicolumn{2}{c|}{$|d_0|/\sigma_{d_0}< 5$  ($< 3$); $|z_0 \sin\theta| < 0.5$~mm}\\[.5ex]
	\cline{2-4}
	& Jets
		&\multicolumn{2}{c|}{$R=0.4$ jets, $\pT>20$~\GeV}\\
	\cline{2-4}
	& \multirow{2}{*}{$b$-jets} 
		& \multirow{2}{*}{\---} &$R=0.4$ jets\\
	&	& & $|\eta|<2.5$, JVT \\
\hline
\hline
\multirow{6}{*}{\emph{Final}} 
	& \multirow{3}{*}{Electrons}
		& $\pT>30$~\GeV & $\pT>65$~\GeV\\
	&	& quality \emph{medium} & quality \emph{tight}\\ 
	&	& \multicolumn{2}{c|}{\emph{loose} isolation}\\
	\cline{2-4}
	& \multirow{3}{*}{Jets} 
		& $R = 0.4$ jets & $R = 1.0$ jets\\
	&	& $\pT>50$~\GeV & $\pT>200$~\GeV \\
	&	& $|\eta|<2.8$, JVT & $|\eta|<2$\\
\hline
\end{tabular}
\end{center}
\end{table}

\begin{table}[htbp]
\caption{Event selection sequences in the \eejj and \enuJ channels. \Wboson-tag50 refers to the $W$-tagger with a 50\% signal efficiency. `Truth matching' requires selected electrons to match electrons from the event generators.}
\label{tab:EvtSelChain}
\begin{center}
\begin{tabular}{|c|c|c|}
\hline
 & \eejj  & \enuJ \\
\hline
\multirow{2}{*}{Overlap removal (1)}
	& between \emph{baseline} & between \emph{baseline}\\
	& electrons, muons, jets & electrons, muons, jets, $b$-jets\\
\hline
Jet cleaning & \multicolumn{2}{c|}{Reject event if it has a \emph{baseline} $R=0.4$ jet 	of non-collision origin} \\
\hline
\multirow{2}{*}{Overlap removal (2)}
	& \multirow{2}{*}{\---} & between \emph{baseline} electrons\\
	& & and \emph{final} $R=1.0$ jets \\[.5ex]
\hline
Number of jets & $N^\mathrm{jets}_\mathrm{final} \ge 2$ & $N^J_\mathrm{final} \ge 1$ \\[.5ex]
\hline
\multirow{2}{*}{Number of leptons}
	&$N^e_\mathrm{final} = 2$ & $N^e_\mathrm{final} = 1$\\
	&$N^e_\mathrm{baseline} \ge 2$ and $N^\mu_\mathrm{baseline} = 0$ & $N^e_\mathrm{baseline} = 1$ and $N^\mu_\mathrm{baseline} = 0$ \\[.5ex]
\hline
Trigger matching & \multicolumn{2}{c|}{ Reject event if \emph{final} electrons are not matched to the trigger objects}\\
\hline
Truth matching & \multicolumn{2}{c|}{ Simulation only: reject event if a selected electron fails truth matching}\\
\hline
\etmiss& \multirow{5}{*}{\---} & $\etmiss > 100$~\GeV \\
\mJ & & $m^J_\mathrm{final}>50$~\GeV \\
\multirow{2}{*}{$D_2^{\beta=1}$} 
	& & Reject event if\\
	& &\emph{final} $R=1.0$ jet does not satisfy\\
	& & upper bound on $D_2^{\beta=1}$ for \Wboson-tag50 \\
\hline
\end{tabular}
\end{center}
\end{table}

One of the background sources common to both channels is a misidentification of hadronic jets, photon conversions in the material or electrons from hadron decays as prompt electrons, referred to as the fake-electron background (\Cref{sec:BackgroundComposition}). 
As this background is estimated in a data-driven way, to avoid double counting, the selected electrons in simulated background events are required to coincide with electrons from the event generators (referred to as `truth matching' in \Cref{tab:EvtSelChain}).
\par
To correct for differences in various object reconstruction and identification efficiencies between the data and simulated event samples, the simulated events are weighted to correct for differences in the trigger, object reconstruction and identification efficiencies between the data and simulation~\cite{Aaboud:2019ynx,Aad:2016jkr,Aaboud:2018xwy}. The correction weights are estimated using measurements in control data samples and are typically consistent with unity to within 5\%.
\par

\section{Background composition}\label{sec:BackgroundComposition}

Background processes in the \eejj final state are dominated by high-mass Drell--Yan $Z/\gamma^* (\rightarrow ee) + \mathrm{jets}$ and $\ttbar \rightarrow bW(e\nu){b}W(e\nu)$ production. Contributions from single-top, diboson, $Z/\gamma^* (\rightarrow \tau\tau) + \mathrm{jets}$, $W (\rightarrow e\nu) + \mathrm{jets}$, and multi-jet production are subdominant. The $W (\rightarrow e\nu) + \mathrm{jets}$ and multi-jet backgrounds contribute to the \eejj sample through misidentification of jets as electrons.\par

The dominant backgrounds in the \enuJ channel are due to the production of a \Wboson boson in association with jets $W (\rightarrow e\nu)+\mathrm{jets}$ and $\ttbar \rightarrow bW(e\nu){b}W(J)$ followed by single-top, $Z/\gamma^* (\rightarrow ee/\tau\tau) +\mathrm{jets}$, $W (\rightarrow \tau\nu) + \mathrm{jets}$, diboson, and multi-jet background production. The only sizeable source of events with a misidentified electron is the multi-jet production.\par

The overall background composition in the \eejj and \enuJ \emph{preselected} event samples is shown in \Cref{tab:yields:pre}.\par

\begin{table}[H]
    \caption{Relative 
	    contributions of background processes to the total number of \emph{preselected} background events.
	    The event yields are normalized to the theoretical cross sections.
	     Contributions included into the fake-electron background are denoted by ``---''.
             The 'fake electron' row includes all sources of events with misidentified electrons.
	     These events are vetoed in the simulated event samples to prevent double counting.
    }
    \label{tab:yields:pre}
    \begin{center}
        \begin{tabular}{|l|c|c|}
        \hline
       &  \eejj $[\%]$ & \enuJ $[\%]$\\ \hline
        $Z/\gamma^* (\rightarrow ee) + \mathrm{jets}$       & \RAIII{79}   & \RAIII{$<1$}  \\
        $Z/\gamma^* (\rightarrow \tau\tau) + \mathrm{jets}$ & \RAIII{$<1$}   & \RAIII{$<1$}  \\
        $W (\rightarrow e\nu) + \mathrm{jets}$              & \RAIII{\---} & \RAIII{27}  \\
        $W (\rightarrow \tau\nu) + \mathrm{jets}$           & \RAIII{\---} & \RAIII{3}   \\
        \ttbar                                              & \RAIII{16}   & \RAIII{58}  \\
        Single-top                                          & \RAIII{1}    & \RAIII{6}   \\
        Fake electron                                       & \RAIII{2}    & \RAIII{2}   \\
        Diboson                                             & \RAIII{2}    & \RAIII{4}   \\ \hline
     \end{tabular}
    \end{center}
\end{table}

Background processes with real electrons are predicted using the simulated event samples. Backgrounds with misidentified electrons are evaluated with a data-driven matrix method as in Ref.~\cite{Aaboud:2016qeg}.\par

\section{Analysis strategy}\label{sec:AnalysisLayout}

The analysis is based on measurements of event yields in a number of phase-space regions defined by the discriminating variables described below. Signal regions (SRs) are constructed to maximize sensitivity to the signal process as predicted by the benchmark model for given values of \mestar, in the presence of the SM background. The signal selection efficiency is nearly independent of $\Lambda$, and therefore the SRs are optimized for the different values of \mestar instead of using a two-dimensional $\Lambda$--\mestar signal optimization. Simulated dominant background processes are constrained in dedicated control regions (CRs). The analysis is blind, and to verify the background predictions after they are constrained by the CRs and before the SRs are unblinded, validation regions (VRs) serve as transitions between CRs and SRs. Signal contamination of all CRs and VRs is negligible. The following section discusses the selection criteria used in the various SRs, CRs, and VRs, which do not overlap.\par

\subsection{Signal regions}
\label{sec:AnalysisLayout:sigreg}

The SRs for the \eejj channel are constructed using the \mll, \st, \mlljj discriminating variables, where

\begin{itemize}
	\item \mll is the invariant mass of the electron pair,
	\item \st is the scalar sum of the transverse momenta of the two electrons and the two jets with the highest \pT, and
	\item \mlljj is the invariant mass of the two electrons and the two jets with the highest \pT.
\end{itemize}

The definition of the SRs is identical to the one used in the search for a singly produced excited muon decaying into a muon and two jets at $\sqrt{s} = 8$~TeV with the ATLAS detector~\cite{Aad:2016giq}.
Further optimization of the \eejj channel SRs does not result in a conclusive improvement of sensitivity to the signal process compared to the initial SR definition given in Ref.~\cite{Aad:2016giq}.
The distributions of the discriminating variables for the \eejj channel are shown in \Cref{Fig:anlayout:eejj} after applying the \emph{preselection} requirements (\Cref{tab:EvtSelChain}) and performing a background-only fit in the corresponding CRs.

\begin{figure}[htbp]
	\centering
	\begin{subfigure}[l]{.48\textwidth}
		\includegraphics[width=1\textwidth]{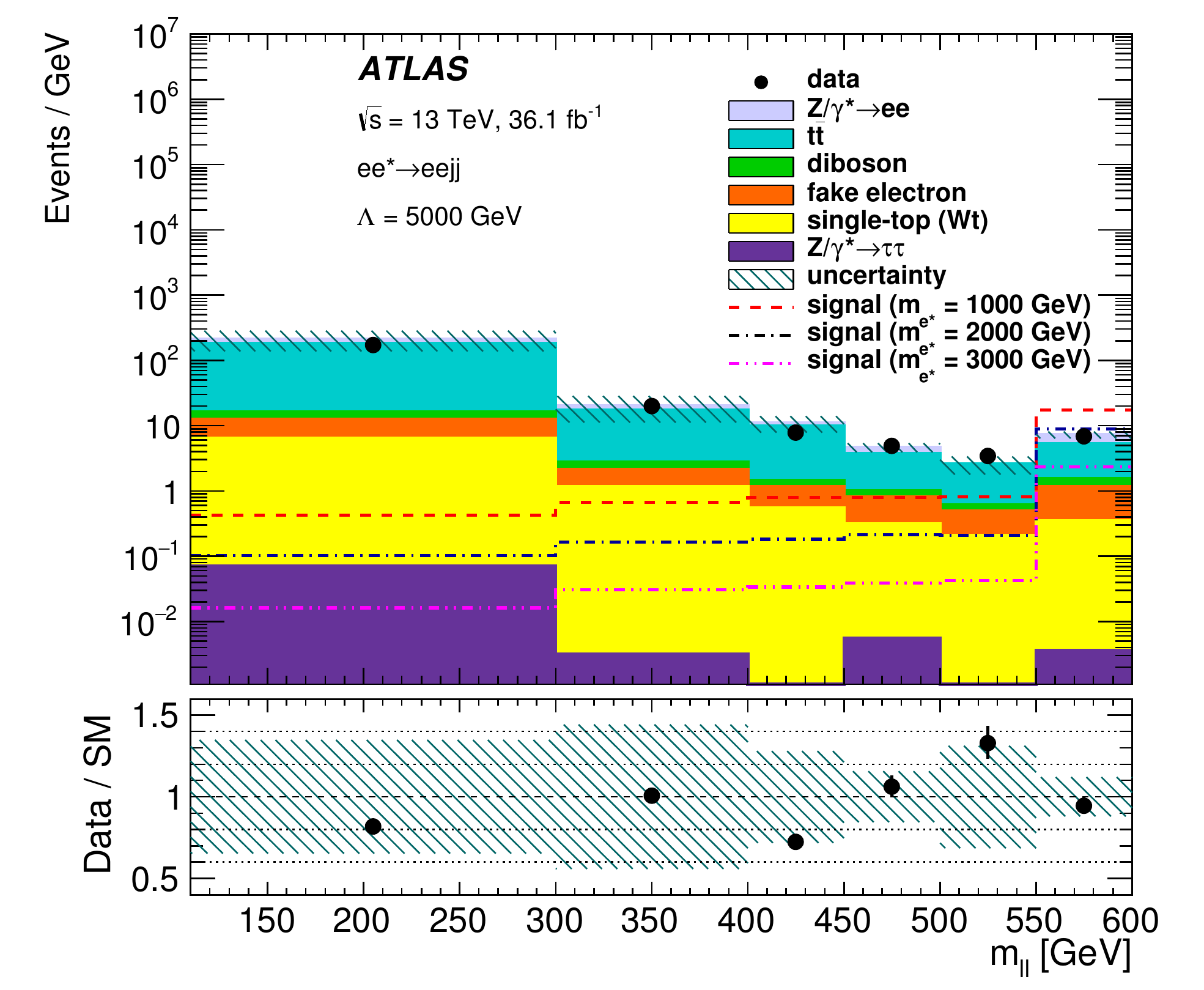}
		\caption{}
		\label{Fig:anlayout:eejj:mll}
	\end{subfigure}
	\hfill
	\begin{subfigure}[r]{.48\textwidth}
		\includegraphics[width=1\textwidth]{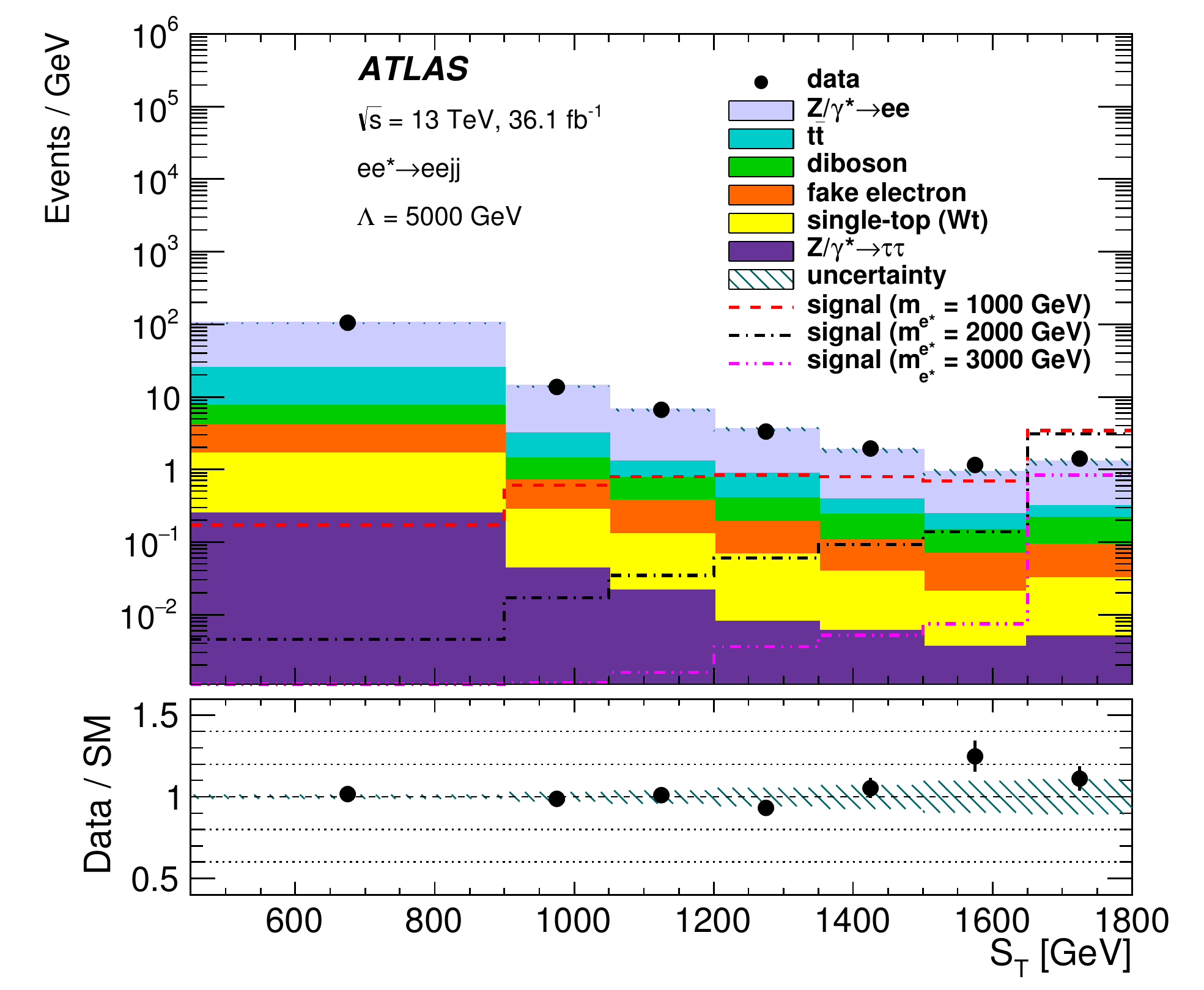}
		\caption{}
		\label{Fig:anlayout:eejj:st}
	\end{subfigure}
	\\
	\begin{subfigure}{.48\textwidth}
		\includegraphics[width=1\textwidth]{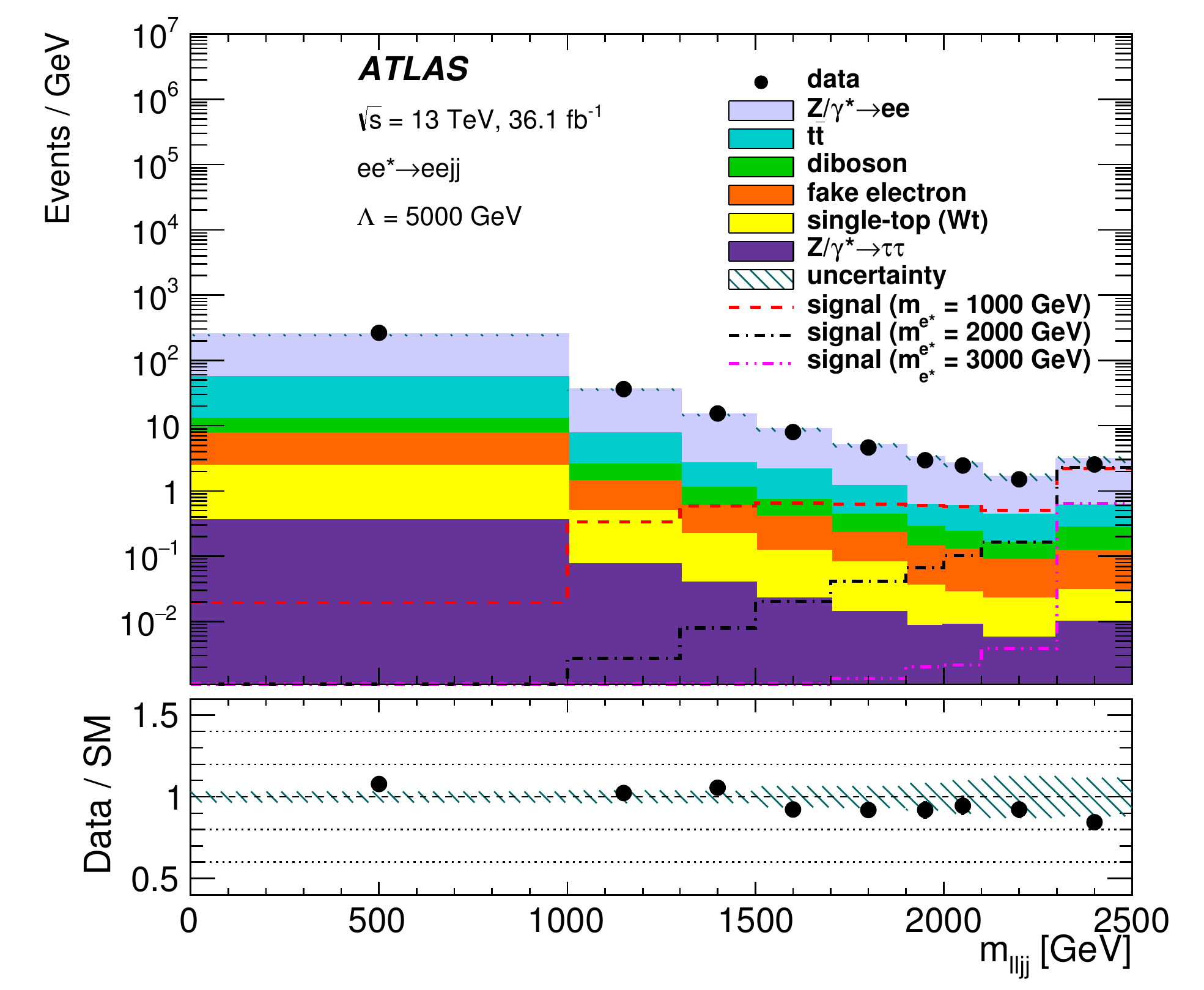}
		\caption{}
		\label{Fig:anlayout:eejj:mlljj}
	\end{subfigure}
	\caption{The distributions of (a) \mll, (b) \st, and (c) \mlljj used to discriminate the signal from background processes in the \eejj channel. The distributions are shown after applying the \emph{preselection} criteria. The background contributions are constrained using the CRs. The signal models assume \Lamb~=~5~\TeV. The last bin includes overflow events (the underflow is not shown). The ratio of the number of data events to the expected number of background events with its statistical uncertainty is shown in the lower panes. The hashed bands represent all considered sources of systematic and statistical uncertainties in the SM background expectation.}
	\label{Fig:anlayout:eejj}
\end{figure}

The selection criteria for the SRs as well as the selection efficiencies for the \eejj channel are shown in \Cref{tab:ELCI_SRDefs}.\par

\begin{table}[htbp]
	\caption{Selection requirements for the SRs used to test various mass hypotheses in the \eejj channel.
	They are applied to the \emph{preselected} event samples (see \Cref{tab:EvtSelChain}). 
	Signal efficiencies are presented as the number of signal events in each SR relative to that after the \emph{preselection} and relative to that before any selection. Each signal region is valid for one or more mass hypotheses, as shown in the second column.}
	\label{tab:ELCI_SRDefs}
	\begin{center}
		\begin{tabular}{|l|c|ccc|c|c|}
			\hline 
			\multirow{2}{*}{}& \mestar & min \mll & min \st & min \mlljj & Efficiency relative to & Total efficiency \\ 
			& $[\mathrm{GeV}]$ & $[\mathrm{GeV}]$ & $[\mathrm{GeV}]$ & $[\mathrm{GeV}]$ & \emph{preselection} stage $[\%]$ & $[\%]$ \\ 
			\hline 
			\multirow{2}{*}{SR1} & \RAIV{100} & \multirow{2}{*}{\RAIV{500}} & \multirow{2}{*}{\RAIV{450}} & \multirow{2}{*}{\RAIV{0}} & \RAII{36} & \RAII{2} \\ 
			 & \RAIV{200} & & & & \RAII{51} & \RAII{10} \\ 
			\hline 
			\multirow{5}{*}{SR2} &\RAIV{300} & \multirow{5}{*}{\RAIV{550}} & \multirow{5}{*}{\RAIV{900}}& \multirow{5}{*}{\RAIV{1000}} & \RAII{41} & \RAII{13}\\ 
			& \RAIV{400} & & & & \RAII{47} & \RAII{18}\\ 
			& \RAIV{500} & & & & \RAII{52} & \RAII{24}\\ 
			& \RAIV{600} & & & & \RAII{57} & \RAII{28}\\ 
			& \RAIV{700} & & & & \RAII{62} & \RAII{33}\\ 
			\hline 
			\multirow{2}{*}{SR3} &\RAIV{800} & \multirow{2}{*}{\RAIV{450}} & \multirow{2}{*}{\RAIV{900}}& \multirow{2}{*}{\RAIV{1300}} & \RAII{68} & \RAII{37}\\ 
			& \RAIV{900} & & & & \RAII{73} & \RAII{41}\\ 
			\hline 
			SR4&1000 &\RAIV{450} &\RAIV{1050} &\RAIV{1300} &  \RAII{73} &  \RAII{43}\\
			\hline                                                  
			SR5&1250 &\RAIV{450} &\RAIV{1200} &\RAIV{1500} &  \RAII{77} &  \RAII{46}\\ 
			\hline                                                  
			SR6&1500 &\RAIV{400} &\RAIV{1200} &\RAIV{1700} &  \RAII{83} &  \RAII{52}\\ 
			\hline                                                  
			SR7&1750 &\RAIV{300} &\RAIV{1350} &\RAIV{1900} &  \RAII{87} &  \RAII{55}\\ 
			\hline                                                  
			SR8&2000 &\RAIV{300} &\RAIV{1350} &\RAIV{2000} &  \RAII{91} &  \RAII{57}\\
			\hline                                                  
			SR9&2250 &\RAIV{300} &\RAIV{1500} &\RAIV{2100} &  \RAII{91} &  \RAII{58}\\
			\hline 
			\multirow{7}{*}{SR10} &\RAIV{2500} & \multirow{7}{*}{\RAIV{110}} & \multirow{7}{*}{\RAIV{1650}}& \multirow{7}{*}{\RAIV{2300}} & \RAII{94} & \RAII{60}\\ 
			& \RAIV{2750} & & & & \RAII{96} & \RAII{61}\\ 
			& \RAIV{3000} & & & & \RAII{97} & \RAII{62}\\ 
			& \RAIV{3250} & & & & \RAII{97} & \RAII{62}\\ 
			& \RAIV{3500} & & & & \RAII{98} & \RAII{62}\\ 
			& \RAIV{3750} & & & & \RAII{98} & \RAII{62}\\ 
			& \RAIV{4000} & & & & \RAII{98} & \RAII{62}\\ 
			\hline 
		\end{tabular}
	\end{center}
\end{table}

The SRs in the \enuJ channel are optimized with discriminating variables at each value of \mestar by maximization of the modified significance defined in Ref.~\cite{TestStatistics} as
\begin{equation}
\label{eq:significance_criterion_GI}
	Z = \sqrt{
			2\times\left(\left(S+B\right)\times\ln\left(1+S/B\right) -S\right)
			},   \nonumber
\end{equation}
where $S$ is the signal yield and $B$ is the background yield in the defined region. This method is checked with minimization of expected upper limit for cross section of the signal, which gives a similar result. 
The maximization is performed by varying the criteria on the set of variables found to provide a maximum discrimination between the signal and the background, \mTnuW and \absdphienu, simultaneously, where:
\begin{itemize}
\item \mTnuW coincides with the transverse mass of the system of the missing transverse momentum and the \Wboson boson in signal events and is given by
\begin{equation}
		\label{eq:mTnuW_GI}
			\mTnuW = \sqrt{
							\left(m_W\right)^2 + 2\times\left(\sqrt{\left(m_W\right)^2 + \left(\pT^W\right)^ 2}\times\etmiss - \px^W\times\exmiss - \py^W\times\eymiss\right) \, ,
							}   \nonumber
\end{equation}
where $p_{x(y)}^W$ is the $x(y)$-component of the momentum of the \Wboson boson candidate reconstructed as the $R = 1.0$ jet.
The \mTnuW is required to exceed a threshold that grows with \mestar.
\item \absdphienu coincides with the absolute value of the azimuthal angle between the neutrino and the electron in signal events.
This quantity provides discrimination between signal events and SM processes involving the leptonic decay of a \Wboson boson.
\end{itemize}

Different sets of selection criteria are examined for each \mestar by applying a maximum or minimum requirement on each of the two variables, i.e.,
min \mTnuW, max \mTnuW, min \absdphienu, and the most effective one is used for the corresponding SR. 
The distributions of the \mTnuW and \absdphienu, as well as in \mJ variables are shown in \Cref{Fig:anlayout:enuJ} for the \enuJ channel after applying the \emph{preselection} requirements and the background-only fit in the CRs as discussed in \Cref{sec:AnalysisLayout:conreg} and \Cref{sec:result}.\par

\begin{figure}[htbp]
	\centering
	\begin{subfigure}[l]{.48\textwidth}
		\includegraphics[width=1\textwidth]{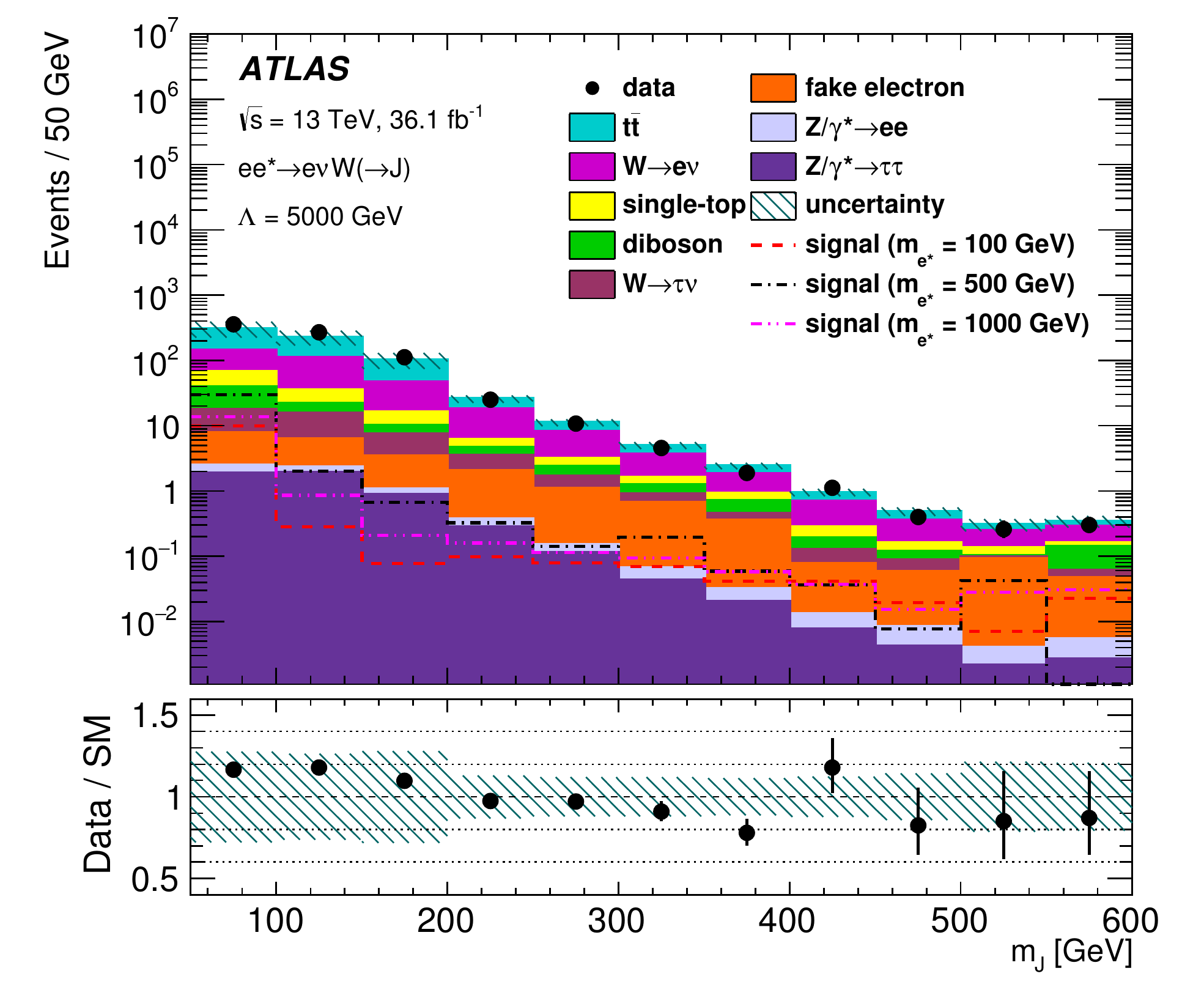}
		\caption{}
		\label{Fig:anlayout:enuJ:mJ}
	\end{subfigure}
	\hfill
	\begin{subfigure}[r]{.48\textwidth}
		\includegraphics[width=1\textwidth]{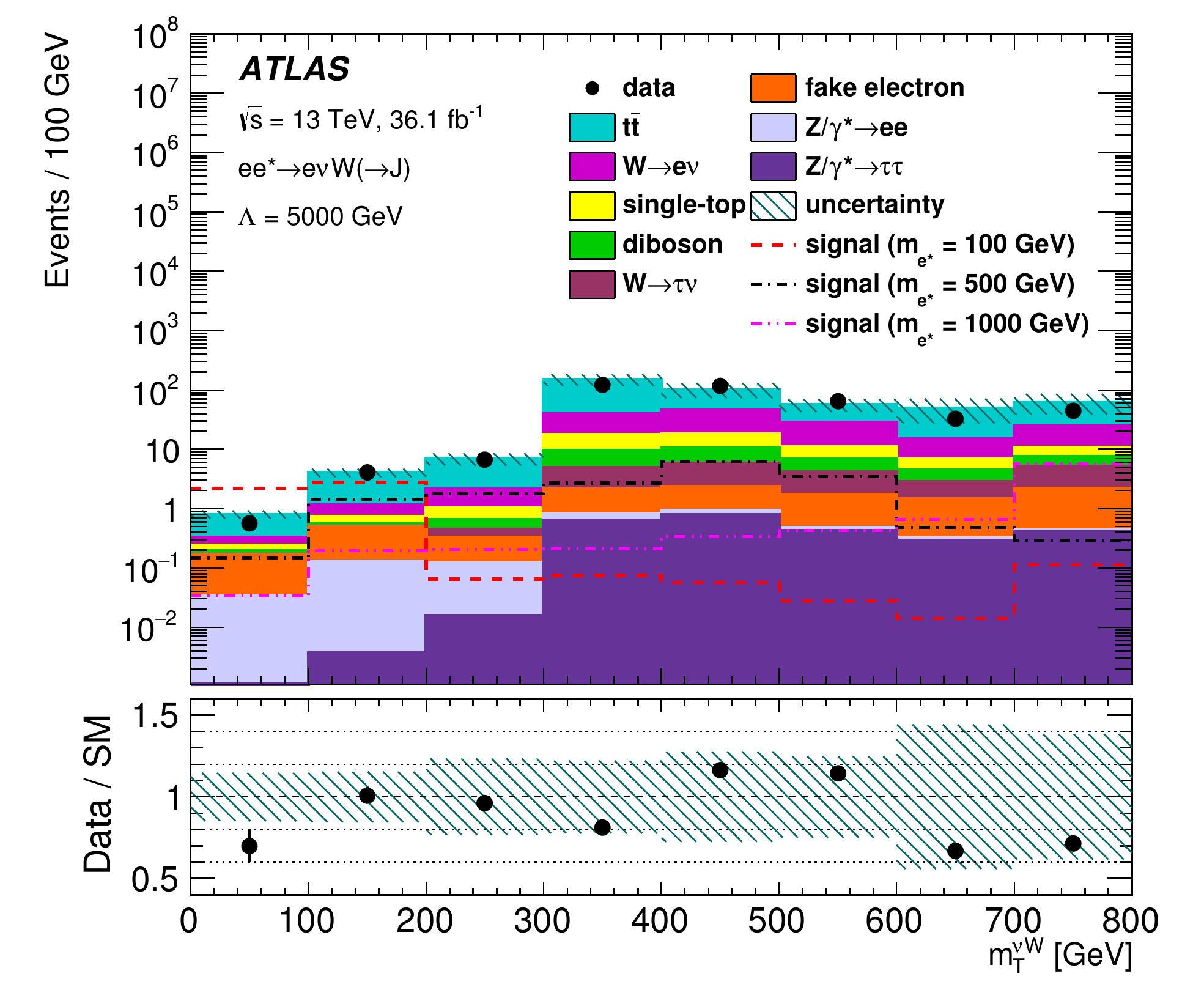}
		\caption{}
		\label{Fig:anlayout:enuJ:mt}
	\end{subfigure}
	\\
	\begin{subfigure}{.48\textwidth}
		\includegraphics[width=1\textwidth]{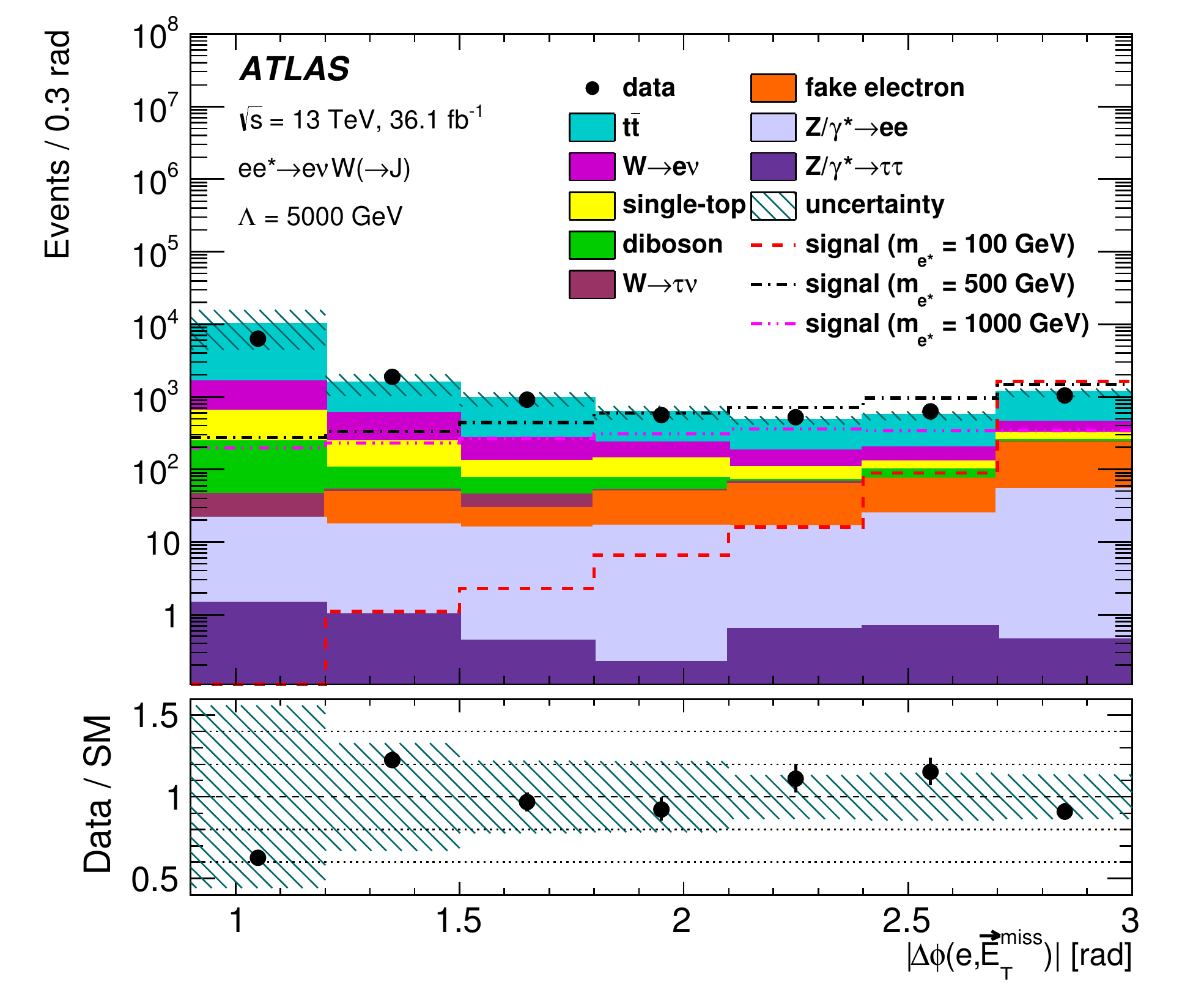}
		\caption{}
		\label{Fig:anlayout:enuJ:dphi}
	\end{subfigure}
	\caption{The distributions of (a) \mJ, (b) \mTnuW, and (c) \absdphienu used to discriminate the signal and background processes in the \enuJ channel. The distributions are shown after applying the \emph{preselection} criteria. The background contributions are constrained using the CRs. The signal models assume \Lamb~=~5~\TeV. The last bin includes overflow events (the underflow is not shown). The ratio of the number of data events to the expected number of background events is shown with its statistical uncertainty in the lower panes. The hashed bands represent all considered sources of systematic and statistical uncertainties for the expected backgrounds.}
	\label{Fig:anlayout:enuJ}
\end{figure}

\begin{table}[htbp]
\caption{Selection requirements for the SRs used to test various mass hypotheses in the \enuJ channel. They are applied to discriminating variables after the \emph{preselection} (\Cref{tab:EvtSelChain}) criteria. Signal efficiencies are presented as the number of signal events in each SR relative to that after the \emph{preselection} and relative to that before any selection. Each signal region is valid for one or more mass hypotheses, as shown in the second column. ``N/A'' means the requirement is not applied for that SR.}
	\label{tab:ELGI_SRDefs}
	\begin{center}
		\begin{tabular}{|l|c|ccc|c|c|}
			\hline 
			\multirow{2}{*}{} & \mestar & min \mTnuW & max \mTnuW & min \absdphienu & Efficiency relative to & Total efficiency \\ 
			& $[\mathrm{GeV}]$ & $[\mathrm{GeV}]$ & $[\mathrm{GeV}]$ & $[\mathrm{radian}]$ & \emph{preselection} stage $[\%]$ & $[\%]$ \\ 
			\hline 
			SR1&  \RAIV{100} &  \RAIII{0}   & \RAIII{200} & 2.7 & 61 &  \RAII{3}\\ 
			\hline 
			SR2&  \RAIV{200} &  \RAIII{100} & \RAIII{N/A} & 2.4 & 59 &  \RAII{4}\\ 
			\hline 
			SR3&  \RAIV{300} &  \RAIII{100} & \RAIII{N/A} & 2.1 & 56 &  \RAII{5}\\ 
			\hline 
			SR4&  \RAIV{400} &  \RAIII{200} & \RAIII{N/A} & 1.8 & 40 &  \RAII{5}\\
			\hline 
			SR5&  \RAIV{500} &  \RAIII{300} & \RAIII{N/A} & 1.5 & 38 &  \RAII{5}\\ 
			\hline 
			SR6&  \RAIV{600} &  \RAIII{400} & \RAIII{N/A} & 1.2 & 38 &  \RAII{6}\\ 
			\hline 
			SR7&  \RAIV{700} &  \RAIII{500} & \RAIII{N/A} & 1.2 & 34 &  \RAII{6}\\ 
			\hline 
			\multirow{2}{*}{SR8}& \RAIV{800} &  \multirow{2}{*}{\RAIII{600}} & \multirow{2}{*}{\RAIII{N/A}} & \multirow{2}{*}{0.9} & 36 & \RAII{7}\\
			& \RAIV{900} &  &  &  & 38 & \RAII{8}\\
			\hline
			\multirow{13}{*}{SR9}& \RAIV{1000} & \multirow{13}{*}{\RAIII{700}} & \multirow{13}{*}{\RAIII{N/A}} & \multirow{13}{*}{0.9} & 37 & \RAII{8}\\
			& \RAIV{1250} & & & & 42 & \RAII{9}\\
			& \RAIV{1500} & & & & 43 & \RAII{10}\\
			& \RAIV{1750} & & & & 44 & \RAII{10}\\
			& \RAIV{2000} & & & & 45 & \RAII{10}\\
			& \RAIV{2250} & & & & 45 & \RAII{10}\\
			& \RAIV{2500} & & & & 43 & \RAII{10}\\
			& \RAIV{2750} & & & & 44 & \RAII{10}\\
			& \RAIV{3000} & & & & 44 & \RAII{10}\\
			& \RAIV{3250} & & & & 42 & \RAII{10}\\
			& \RAIV{3500} & & & & 43 & \RAII{10}\\
			& \RAIV{3750} & & & & 42 & \RAII{10}\\
			& \RAIV{4000} & & & & 42 & \RAII{9}\\
			\hline
		\end{tabular}
	\end{center}
\end{table}

In the \enuJ channel, the observables \mTnuW and \absdphienu are used to create the nine optimized SRs.

Each SR targets a model with a given mass of the excited electron.  
The SR is defined by applying the  \emph{preselection} introduced in~\Cref{sec:EventSelection} and additionally requiring the criteria defined in \Cref{tab:ELGI_SRDefs}, which include a $b$-jet veto and selection on \mTnuW and \absdphienu. The large-$R$ jet also passes the 50\% signal efficiency requirement on \mJ from the $W$-tagger.\par

\subsection{Control regions}
\label{sec:AnalysisLayout:conreg}

The control regions are used to derive normalization factors and to constrain systematic uncertainties in the respective background yields (\Cref{sec:result}). The CRs are defined so as to ensure a high purity in the corresponding background processes and a sufficient number of events, while having no overlap with events in the respective SRs.  To ensure that extrapolation uncertainties are small, the selection criteria for the CRs closely follow those used in the corresponding SRs. An individual selection criterion is changed to enrich the background of interest while ensuring no overlap with the signal region. Hence, separate control regions are defined for each signal region. The other selection criteria are the same as for the signal regions.\par

The CRs of the \eejj channel (\Cref{tab:ELCI_CRVRDefs}) are introduced for the two largest sources of background, \zjets and \ttbar. The \zgamma CRs are defined by requiring \mllzwindows~< 20~\GeV\ and the same \st and \mlljj selections as in the corresponding SRs. The \ttbar CRs are defined by the full SR selections but at the \emph{preselection} require a single-muon trigger and exactly one electron and exactly one muon in the event, leading to an \emjj signature. The kinematic criteria used for the \emjj signature (apart from the lepton preselection) are identical to those in the nominal \eejj SR selection.\par

The CRs for the \enuJ channel (\Cref{tab:ELGI_CRSRDefs}) are defined for the \wjets and \ttbar background processes. The \Wboson CR is defined by applying the same selection requirements as in the SRs (\Cref{tab:ELGI_SRDefs}), including the $b$-jet veto, but requiring the jets to fail the boosted jet mass $W$-tagger with the 80\% signal efficiency ($W$-tag80). Also, the \absdphienu selection is removed for all \Wboson CRs in order to reduce the statistical uncertainties. There is no \Wboson CR corresponding to SR1 since the \wjets background process is subdominant in such a CR. The \ttbar CR events are required to have at least two $b$-jets, fulfil the respective SR selections from \Cref{tab:ELGI_SRDefs}, and have a leading large-$R$ jet satisfying the \mJ $W$-tag50 criterion. No additional requirements on the kinematic properties of the $b$-jets are applied in the \ttbar CR. The \ttbar background prediction is corrected for the difference in $b$-jet identification efficiencies between data and simulated events, and the corresponding systematic uncertainties are accounted for. 
Theoretical uncertainties in the \ttbar kinematic distributions are accounted for as described in \Cref{sec:SystematicUncertainties}.
\par

\subsection{Validation regions}
\label{sec:AnalysisLayout:valreg}

The background estimation in the CRs is validated in additional phase space regions, the VRs. The VRs are not included in any fits aimed at a signal search. \par

In the \eejj channel, a \mll VR is defined as the intermediate range between SR and \zgamma CR. A further requirement on the \etmiss is introduced to split the \mll VR into regions dominated by \zgamma and \ttbar processes. A same-sign (SS) VR is defined in order to validate the fake-electron background estimate by selecting events with \mll~>~160~\GeV\ in which both electrons are required to have the same electric charge $Q_e$ (\Cref{tab:ELCI_CRVRDefs}).\par

The \mJ and $b$-jet VRs are introduced for the \enuJ channel. The \mJ VRs are defined by applying the \emph{preselection} requirements while inverting the requirement on the boosted jet mass \Wboson-tagger interval relative to the \Wboson CRs and SRs (\Cref{tab:ELGI_CRSRDefs}). The $b$-jet VRs require the number of $b$-jets to be equal to one to validate the application of \ttbar normalization derived in \ttbar CR with the two $b$-jets requirement to the SR with zero $b$-jets. The requirements on \mTnuW and \absdphienu in the VRs are the same as in the corresponding SRs.\par

\begin{table}[htbp]
	\caption{Selection requirements applied in addition to the \emph{preselection} (\Cref{tab:EvtSelChain}) in the CRs, VRs, and SRs for the \eejj channel. 
	         ``Pass'', `fail'' or ``N/A'' mean that the requirement is passed, failed or not applied, respectively.
	}
	\label{tab:ELCI_CRVRDefs}
	\begin{center}
		\renewcommand{\arraystretch}{1.5}
		\begin{tabular}{|c|c|c|c|c|c|}
			\hline 
			Region & Leptons & \mll & \st & \mlljj & $Q_e$\\
			\hline 
			SR 			& 2 electrons & pass & pass & pass & N/A\\
			\hline
			\multirow{2}{*}{\zgamma CR} & \multirow{2}{*}{2 electrons} & $>70$~\GeV and & \multirow{2}{*}{pass} & \multirow{2}{*}{pass} & \multirow{2}{*}{N/A}\\
			& & $<110$~\GeV & & & \\
			\hline 
			\multirow{2}{*}{\ttbar CR} & 1 electron & \multirow{2}{*}{pass} & \multirow{2}{*}{pass} & \multirow{2}{*}{pass} & \multirow{2}{*}{N/A}\\
			& and 1 muon & & & & \\
			\hline
			\multirow{2}{*}{\mll VR} & \multirow{2}{*}{2 electrons} & $>110$~GeV and & \multirow{2}{*}{pass} & \multirow{2}{*}{pass} & \multirow{2}{*}{N/A}\\
			& & $<\mll^{\mathrm{SR\ threshold}}$ & & & \\
			\hline 
			SS VR 		& 2 electrons & $>160$~GeV & N/A & N/A & $Q_{e1} = Q_{e2}$\\
			\hline 
		\end{tabular}
	\end{center}
\end{table}

\begin{table}[htbp]
	\caption{Selection requirements applied in addition to the \emph{preselection} (\Cref{tab:EvtSelChain}) in the CRs, VRs and SRs for the \enuJ channel. 
	         ``Pass'', `fail'' or ``N/A'' mean that the requirement is passed, failed or not applied, respectively.
		 $W$-tag80 refers to the working point of the $W$-tagger with 80\% signal efficiency.
	}
	\label{tab:ELGI_CRSRDefs}
	\begin{center}
		\begin{tabular}{|c|c|c|c|c|}
			\hline 
			Region & \mJ interval & $N^{b\textrm{-jets}}$ & \mTnuW & \absdphienu \\
			\hline
			\multirow{2}{*}{SR} & $W$-tag50 & \multirow{2}{*}{0} & \multirow{2}{*}{pass} & \multirow{2}{*}{pass}\\
			& pass & & & \\			
			\hline
			\multirow{2}{*}{\Wboson CR} & $W$-tag80 & \multirow{2}{*}{0} & \multirow{2}{*}{pass} & \multirow{2}{*}{N/A}\\
			& fail & & & \\			
			\hline 
			\multirow{2}{*}{\ttbar CR} & $W$-tag50 & \multirow{2}{*}{$\geq 2$} & \multirow{2}{*}{pass} & \multirow{2}{*}{pass}\\
			& pass & & & \\			
			\hline
			\multirow{2}{*}{\mJ VR} & $W$-tag50 fail & \multirow{2}{*}{N/A} & \multirow{2}{*}{pass} & \multirow{2}{*}{pass}\\
			& $W$-tag80 pass & & & \\			
			\hline
			\multirow{2}{*}{$b$-jet VR} & $W$-tag50 & \multirow{2}{*}{1} & \multirow{2}{*}{pass} & \multirow{2}{*}{pass}\\
			& pass & & & \\			
			\hline
		\end{tabular}
	\end{center}
\end{table}

\section{Systematic uncertainties}\label{sec:SystematicUncertainties}

The systematic uncertainties of the search are divided into two categories:
the experimental uncertatinties and theoretical uncertainties in signal and background prediction.
Details of the evaluation of experimental uncertainties are provided in the references in \Cref{sec:EventSelection}.
\par
The uncertainty in the combined 2015+2016 integrated luminosity is 2.1\%. 
It is derived from the calibration of the luminosity scale using $x$-$y$ beam-separation scans, 
following a methodology similar to that detailed in Ref.~\cite{Aaboud:2016hhf}, and using the LUCID-2 detector for the baseline luminosity measurements \cite{LUCID2}.
\par
The uncertainties in the electron energy scale and resolution result in less than a 1\% effect for simulated background or signal event yields in the SRs. In addition, uncertainties are taken into account for the electron trigger ($< 2\%$), identification ($< 3\%$), and reconstruction ($< 1\%$) efficiencies, and for the isolation requirements ($< 6\%$).
\par
The effect of the uncertainty in the muon momentum on \ttbar background event yields in the \ttbar CRs of the \eejj channel does not exceed 1\%. Differences between data and simulated event samples in the muon identification and trigger efficiencies are also taken into account and are less than 1\%.
\par
The impact of the $R=0.4$ jet energy scale (JES) and resolution (JER) uncertainties on the background event yields is 1--5\% (JES) and 1--6\% (JER) in the SRs of the \eejj channel. 
The signal selection efficiency change in the SRs of the \eejj channel due to the JES uncertainties never exceeds 2\%, while the effect of the JER uncertainty is negligible. Uncertainties associated with $R=1.0$ jets in the \enuJ channel arise from uncertainties in the calibration of the JES and the jet mass scale.
The impact on the background event yields in the SRs ranges between 20\% and 40\%, and the effect on signal yields is below 10\%. Uncertainties related to the $b$-tagging efficiency corrections are also taken into account in the \enuJ channel, and the effect on \ttbar yields is always below 5\%.
\par
The procedure to estimate fake-electron background includes a systematic uncertainty, which is 10--40\% of the fake-electron background estimate in the SRs,  depending on the \pT of the electron candidates.
\par
Theoretical uncertainties affect the simulated event samples of backgrounds and signal. 
For the background samples, they lie in the PDF set, the value of $\alpha_S$, and missing higher order corrections in perturbative calculations. 
The latter effect is estimated by varying the renormalization and factorization scales by factors of one-half and two, excluding those variations where both differ by factor of four.
The PDF uncertainty is estimated using the envelope of the NNPDF3.0 PDF set~\cite{Botje:2011sn} and the two alternative PDF sets,
the MMHT2014~\cite{Harland-Lang:2014zoa} and CT14nnlo~\cite{Dulat:2015mca}. 
The uncertainty due to $\alpha_S$ is estimated by varying its nominal value of 0.118 by $\pm$0.001.
For the \ttbar background, the theoretical uncertainty also includes 
effects of the matching between ME and PS via the variation of the \POWHEGBOX $h_{\mathrm{damp}}$ parameter.
The effects of the ME and hadronization model choice are assessed for \ttbar and single-top MC samples by 
replacing the \POWHEGBOX ME by aMC@NLO~\cite{Frixione:2010wd} and the \PYTHIA~8 hadronization model by the one implemented in \HERWIG~7~\cite{Bahr:2008pv}.
The theoretical uncertainties in the signal prediction are estimated using the PDF set variations only.
The theoretical uncertainties for background yields range from 7\% to 22\% in the SRs of the \eejj channel and from 3\% to 10\% in the SRs of the \enuJ channel.\par

\section{Statistical analysis and results}\label{sec:result}

The statistical analysis of the search is based on a maximum-likelihood fit. The signal hypothesis test is performed using a likelihood-ratio test statistic in the asymptotic approach~\cite{TestStatistics}.
\par
The likelihood function is constructed as the product of Poisson probabilities of the SR and the CRs as in Ref.~\cite{Cranmer:2012sba}. 
The normalizations of the backgrounds which have CRs, i.e., \zjets and \ttbar for the \eejj channel and \wjets and \ttbar for the \enuJ channel, are free parameters, denoted by $\beta$, in the fit. 
These corrections are used to scale the background predictions in the SRs.
Their values and uncertainties after the background-only fit in the CRs are summarized in \Cref{tab:beta}. 
The deviation of the $\beta$ values from unity reflects the fact that at high \st the simulated events do not accurately describe the data. This is also observed, for example, in the leptoquark search by ATLAS~\cite{Aaboud:2019jcc}. 
After the fit in the corresponding CR, the background yields agree with the data in all VRs within the uncertainties.
The final fit combining CRs and SRs results in negligible shifts of the background normalization factors with respect to the CR-only fits.
Systematic uncertainties are incorporated into the likelihood function with a set of nuisance parameters with Gaussian constraint terms. 
Statistical uncertainties from the simulated event samples are included as nuisance parameters with Poisson constraint terms. 
Correlations of the systematic uncertainty effects across regions are taken into account. 
The signal normalization (strength) is obtained by maximizing the likelihood function for each signal hypothesis. 
The statistical analysis is performed using the RooStats~\cite{RooStats} and HistFitter~\cite{HistFitter} software.\par

\begin{table}[htbp]
	\caption{
Background normalization factors with 68\% confidence intervals after the background-only fit in the CRs. 
CRs not defined in the \enuJ channel are denoted as ``N/A''.
The $\beta_{W}$ normalization factor in the \enuJ SR1 is fixed to unity.
	        }
	\label{tab:beta}
	\begin{center}
		\begin{tabular}{|l|c|c|c|c|}
			\hline
			\multirow{2}{*}{} &\multicolumn{2}{c|}{\eejj} &\multicolumn{2}{c|}{\enuJ} \\
			\cline{2-5}
			& $\beta_{\zgamma}$ & $\beta_{\ttbar}$ & $\beta_{W}$ & $\beta_{\ttbar}$ \\
			\hline 
			CR1 & $0.94^{+0.04}_{-0.04}$ & $0.95^{+0.08}_{-0.07}$ & N/A                                & $0.8^{+0.2}_{-0.2}$ \\
			CR2 & $0.82^{+0.04}_{-0.04}$ & $1.0^{+0.2}_{-0.2}$      & $0.79^{+0.08}_{-0.08}$ & $0.8^{+0.2}_{-0.2}$ \\
			CR3 & $0.79^{+0.04}_{-0.04}$ & $0.8^{+0.2}_{-0.2}$      & $0.79^{+0.08}_{-0.08}$ & $0.8^{+0.2}_{-0.2}$ \\ 
			CR4 & $0.81^{+0.05}_{-0.05}$ & $0.8^{+0.3}_{-0.3}$      & $0.77^{+0.10}_{-0.10}$ & $1.0^{+0.4}_{-0.3}$ \\		
			CR5 & $0.80^{+0.06}_{-0.05}$ & $1.3^{+0.5}_{-0.4}$      & $0.72^{+0.10}_{-0.10}$ & $1.2^{+0.5}_{-0.4}$ \\
			CR6 & $0.76^{+0.06}_{-0.06}$ & $1.4^{+0.5}_{-0.5}$      & $0.83^{+0.10}_{-0.10}$ & $0.7^{+0.4}_{-0.4}$ \\
			CR7 & $0.78^{+0.07}_{-0.07}$ & $1.0^{+0.6}_{-0.5}$      & $0.91^{+0.11}_{-0.18}$ & $0.13^{+1.17}_{-0.13}$ \\
			CR8 & $0.74^{+0.07}_{-0.07}$ & $1.2^{+0.8}_{-0.7}$      & $0.65^{+0.15}_{-0.22}$ & $1.7^{+1.6}_{-0.9}$ \\
			CR9 & $0.64^{+0.08}_{-0.07}$ & $1.4^{+1.1}_{-0.9}$      & $0.66^{+0.14}_{-0.20}$ & $1.6^{+1.6}_{-0.9}$ \\
			CR10 & $0.62^{+0.10}_{-0.09}$ & $1.3^{+0.7}_{-0.5}$    & N/A                                 & N/A \\
			\hline 
		\end{tabular}
	\end{center}
\end{table}

The observed and expected yields in the SRs for the \eejj and \enuJ channels after the combined maximum-likelihood fits to only background processes in the CRs and SRs are shown in \Cref{yields:table:SR:CI,yields:table:SR:GI}, respectively.
When calculating the uncertainties on the expected yields in the SRs, all correlations between the nuisance parameters estimates are taken into account. 
No significant excess above the expected SM background is observed, and limits on the excited lepton model parameters are set at 95\% confidence level (CL), using the CL$_\mathrm{s}$ method~\cite{CLs}. The upper limits on the signal production cross section times branching ratio $\sigma \times \mathcal{B}$ as a function of \mestar are presented in \Cref{fig:eejj:limits:xs,fig:enuJ:limits:xs} for the \eejj and \enuJ channels, respectively.
The fluctuations observed in the limit for the \enuJ channel for \mestar points below 1~\TeV\ are caused by the selection criteria optimized separately at each mass points.
\par
\begin{sidewaystable}
	\caption{Event yields in the signal regions of the \eejj channel. 
	         Data are shown together with the background contributions after a combined background-only fit in the control and signal regions.
		}
	\label{yields:table:SR:CI}
	\begin{center}
	\setlength{\tabcolsep}{0.2pc}
		{
		\small
			\begin{tabular}{l|c|c|c|c|c|c|c|c|c|c}
				\hline
				\textbf{Yields}                  & SR1           & SR2           & SR3           & SR4           & SR5             & SR6             & SR7             & SR8             & SR9             & SR10  \\
				\hline
Observed & $448$ & $102$ & $94$ & $62$ & $37$ & $32$ & $32$ & $27$ & $16$ & $15$ \\
\hline
Background & \CPM{430}{50} & \CPM{110}{15} & \CPM{96}{10} & \CPM{71}{8} & \CPM{43}{6} & \CPM{40}{5} & \CPM{29}{5} & \CPM{27}{5} & \CPM{18}{3} & \CPM{21}{3} \\
\hline
$Z/\gamma^*\rightarrow ee$ & \CPM{100}{20} & \CPM{48}{10} & \CPM{37}{6} & \CPM{30}{5} & \CPM{16}{2} & \CPM{14}{2} & \CPM{10}{2} & \CPM{9}{2} & \CPM{5.4}{1.1} & \CPM{8.0}{1.3} \\
$t\bar{t}$ & \CPM{230}{30} & \CPM{33}{7} & \CPM{24}{6} & \CPM{14}{5} & \CPM{12}{4} & \CPM{11}{4} & \CPM{6}{4} & \CPM{7}{4} & \CPM{4}{3} & \CPM{5}{2} \\
Single-top $(Wt)$ & \CPM{24}{3} & \CPM{7.8}{0.8} & \CPM{9.2}{1.0} & \CPM{7.3}{0.7} & \CPM{4.1}{0.4} & \CPM{4.6}{0.4} & \CPM{3.0}{0.3} & \CPM{2.4}{0.3} & \CPM{1.6}{0.2} & \CPM{1.50}{0.14} \\
Fake electron & \CPM{50}{9} & \CPM{14}{2} & \CPM{18}{2} & \CPM{15}{2} & \CPM{7.2}{0.8} & \CPM{7.5}{0.9} & \CPM{6.9}{0.7} & \CPM{6.7}{0.6} & \CPM{4.9}{0.5} & \CPM{4.7}{0.5} \\
Diboson & \CPM{21}{7} & \CPM{6}{2} & \CPM{8}{2} & \CPM{6}{2} & \CPM{3.8}{1.1} & \CPM{3.6}{1.0} & \CPM{3.0}{0.8} & \CPM{2.6}{0.7} & \CPM{1.8}{0.5} & \CPM{2.5}{0.7} \\
$Z/\gamma^*\rightarrow \tau\tau$ & \CPM{0.22}{0.09} & \CPM{0.12}{0.04} & \CPM{0.12}{0.04} & \CPM{0.04}{0.01} & \CPM{0.03}{0.01} & \CPM{0.03}{0.01} & \CPM{0.03}{0.01} & \CPM{0.03}{0.01} & \CPM{0.01}{0.01} & \CPM{0.04}{0.02} \\
\hline
			\end{tabular}
		}
	\end{center}
\end{sidewaystable}

\begin{sidewaystable}
	\caption{Event yields in the signal regions of the \enuJ channel. 
	         Data are shown together with the background contributions after a combined background-only fit in the control and signal regions.
		}
	\label{yields:table:SR:GI}
	\begin{center}
	\setlength{\tabcolsep}{0.2pc}
		{
		\small
			\begin{tabular}{l|c|c|c|c|c|c|c|c|c}
				\hline
				\textbf{Yields} & SR1 & SR2 & SR3 & SR4 & SR5 & SR6 & SR7 & SR8 & SR9 \\
				\hline
Observed & $13$ & $25$ & $39$ & $35$ & $43$ & $34$ & $15$ & $16$ & $8$ \\
\hline
Background & \CPM{13}{5} & \CPM{17}{5} & \CPM{26}{8} & \CPM{25}{5} & \CPM{34}{8} & \CPM{30}{6} & \CPM{12}{4} & \CPM{8}{2} & \CPM{6}{2} \\
\hline
$W\rightarrow e\nu$ & \CPM{2}{2} & \CPM{7}{3} & \CPM{11}{4} & \CPM{13}{3} & \CPM{14}{5} & \CPM{17}{4} & \CPM{7}{3} & \CPM{3.2}{1.3} & \CPM{2.2}{1.1} \\
$Z/\gamma^*\rightarrow ee$ & \CPM{1.3}{1.2} & \CPM{1.6}{1.1} & \CPM{2.1}{1.3} & \CPM{1.7}{1.0} & \CPM{1.4}{0.9} & \CPM{0.6}{0.3} & \CPM{0.14}{0.10} & \CPM{0.10}{0.05} & \CPM{0.04}{0.03} \\
\ttbar & \CPM{2.9}{1.2} & \CPM{5}{2} & \CPM{7}{3} & \CPM{4}{3} & \CPM{11}{5} & \CPM{3}{2} & \CPM{0.1}{0.3} & \CPM{1}{2} & \CPM{0.4}{0.7} \\
Single-top & \CPM{0.7}{0.3} & \CPM{1.9}{0.5} & \CPM{2.6}{0.6} & \CPM{3.0}{1.4} & \CPM{3.0}{1.4} & \CPM{4}{2} & \CPM{1.9}{0.7} & \CPM{1.9}{0.6} & \CPM{1.7}{0.7} \\
Fake electron & \CPM{6}{2} & \CPM{1.9}{0.3} & \CPM{3.2}{1.0} & \CPM{0.6}{0.3} & \CPM{0.25}{0.07} & \CPM{0.7}{0.3} & \CPM{0.06}{0.11} & \--- & \--- \\
Diboson & \CPM{0.0}{1.1} & \CPM{0.2}{1.1} & \CPM{1}{2} & \CPM{2.9}{0.9} & \CPM{3.4}{0.7} & \CPM{3}{2} & \CPM{2}{3} & \CPM{1.1}{1.1} & \CPM{1.1}{1.1} \\
$W\rightarrow \tau\nu$ & \--- & \--- & \CPM{0.0}{0.5} & \CPM{0.1}{0.2} & \CPM{0.34}{0.10} & \CPM{0.2}{0.6} & \CPM{0.21}{0.09} & \CPM{0.21}{0.08} & \CPM{0.20}{0.09} \\
$Z/\gamma^*\rightarrow \tau\tau$ & \CPM{0.04}{0.02} & \CPM{0.04}{0.02} & \CPM{0.06}{0.03} & \CPM{0.03}{0.02} & \CPM{0.08}{0.06} & \CPM{0.06}{0.05} & \CPM{0.05}{0.06} & \CPM{0.1}{0.2} & \--- \\
				\hline
			\end{tabular}
		}
	\end{center}
\end{sidewaystable}

\begin{figure}[htbp]
	\begin{subfigure}[t]{.48\textwidth}
	\centering
		\includegraphics[width=1\textwidth]{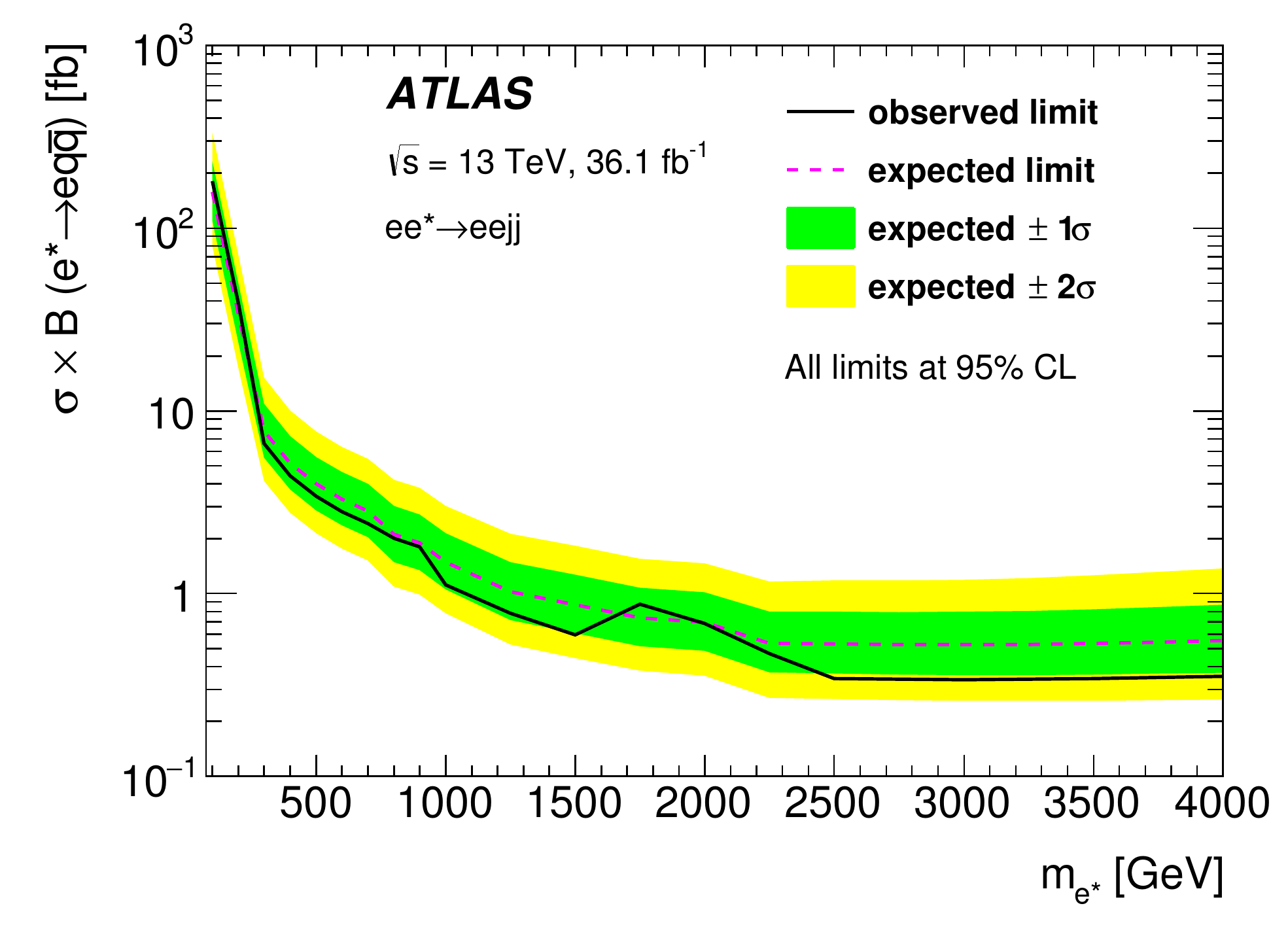}
		\caption{}
		\label{fig:eejj:limits:xs}
	\end{subfigure}
	\hfill
	\begin{subfigure}[t]{.48\textwidth}
	\centering
		\includegraphics[width=1\textwidth]{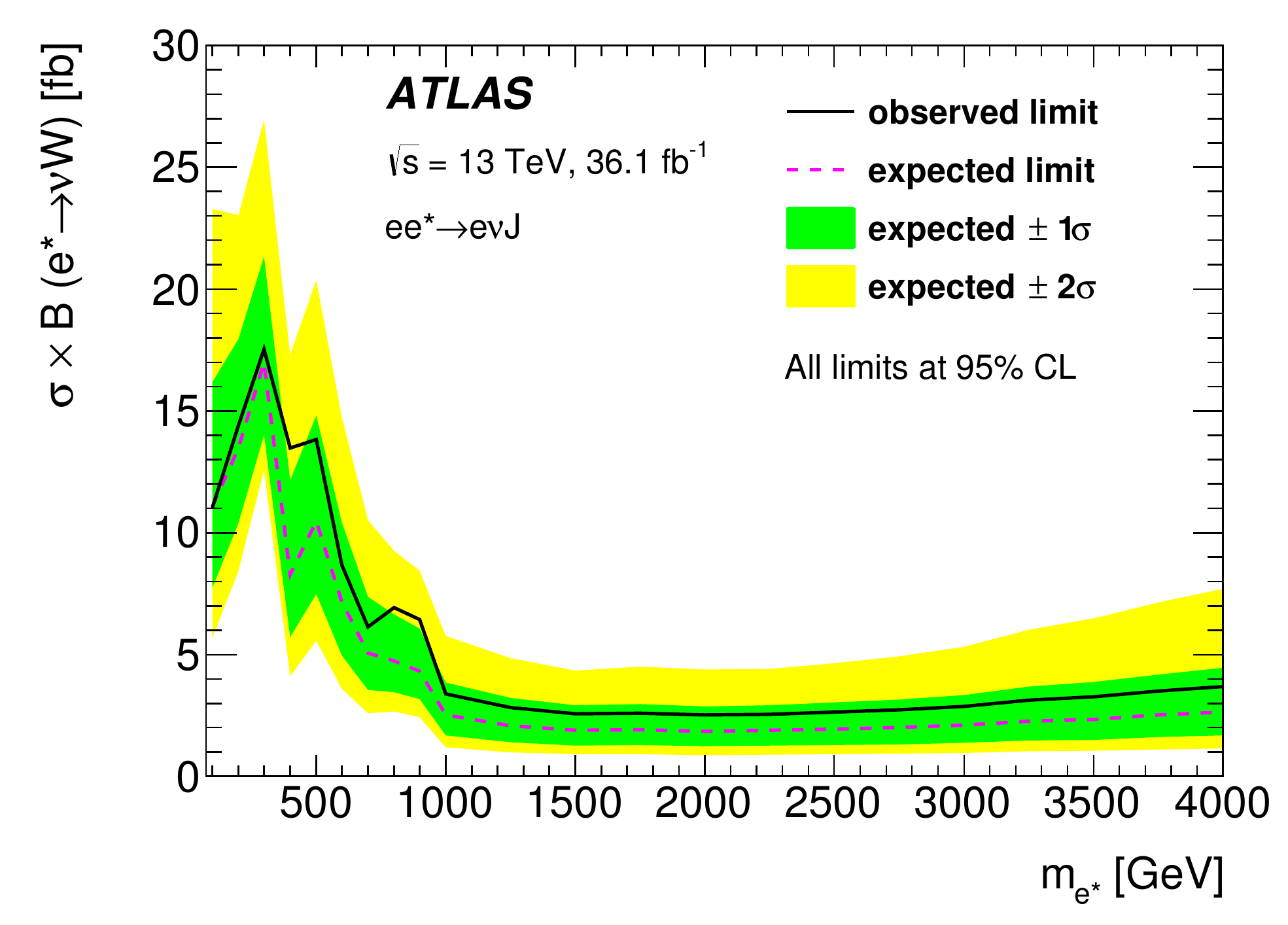}
		\caption{}
		\label{fig:enuJ:limits:xs}
	\end{subfigure}
	\caption{Upper limits on $\sigma \times \mathcal{B}$ as a function of \mestar in (a) the \eejj channel and (b) the \enuJ channel.
	The $\pm1(2)\sigma$ uncertainty bands around the expected limit represent all sources of systematic and statistical uncertainties.}
\end{figure}

The lower limits on the compositeness scale parameter $\Lambda$ as a function of \mestar for the \eejj and \enuJ channels are presented in \Cref{fig:eejj:limits:la,fig:enuJ:limits:la}. They are calculated from the upper limits on $\sigma \times \mathcal{B}$, taking into account the $\mathcal{B}$ dependency on both the \mestar and $\Lambda$ parameters. The limits on $\Lambda$ in the \eejj channel are extrapolated to the values of $\mestar > 4$~\TeV, since the signal selection efficiency remains constant for the highest \mestar values in SR10, as is shown in \Cref{tab:ELCI_SRDefs}. A unified likelihood function is constructed for the \eejj and \enuJ channels at each \mestar value considered in order to extract a combined limit on \Lamb as a function of \mestar. The correlations of systematic uncertainty effects between the two search channels are included. The combined limit is presented in \Cref{fig:comb:limits:la} along with the individual limits from the \eejj and \enuJ channels as well as the limit set by ATLAS in the $ee\gamma$ search channel at \rts~=~8~\TeV~\cite{Aad:2013jja}.

Observed and expected model-independent upper limits on the number of signal events in the signal regions of the \eejj and \enuJ channels are shown in \Cref{tab:modindlim} along with the upper limits on the visible signal cross section, which is defined as the production cross-section times the overall signal efficiency.

\begin{figure}[htbp]
	\begin{subfigure}[t]{.48\textwidth}
	\centering
		\includegraphics[width=1\textwidth]{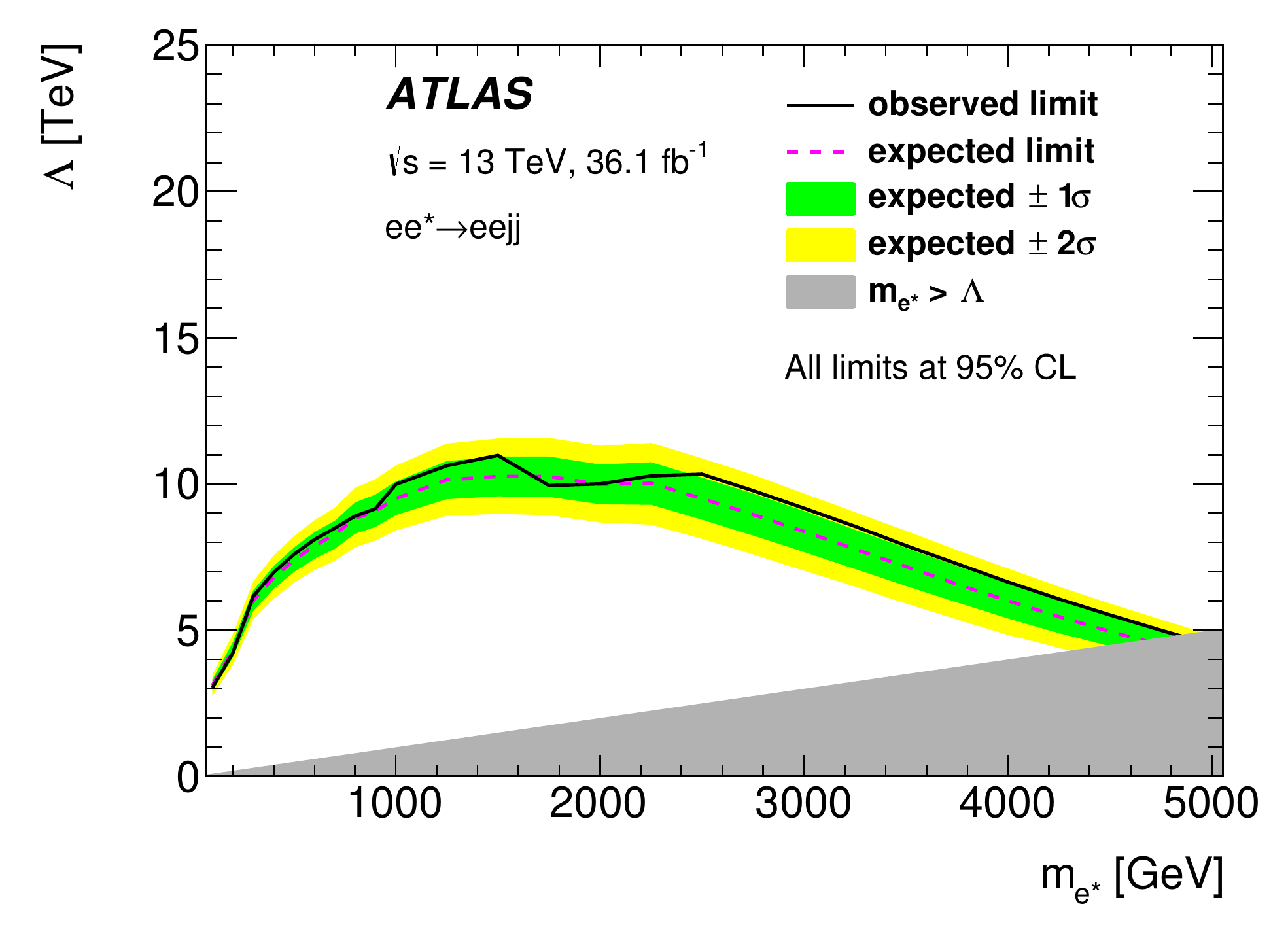}
		\caption{}
		\label{fig:eejj:limits:la}
	\end{subfigure}
	\hfill
	\begin{subfigure}[t]{.48\textwidth}
	\centering
		\includegraphics[width=1\textwidth]{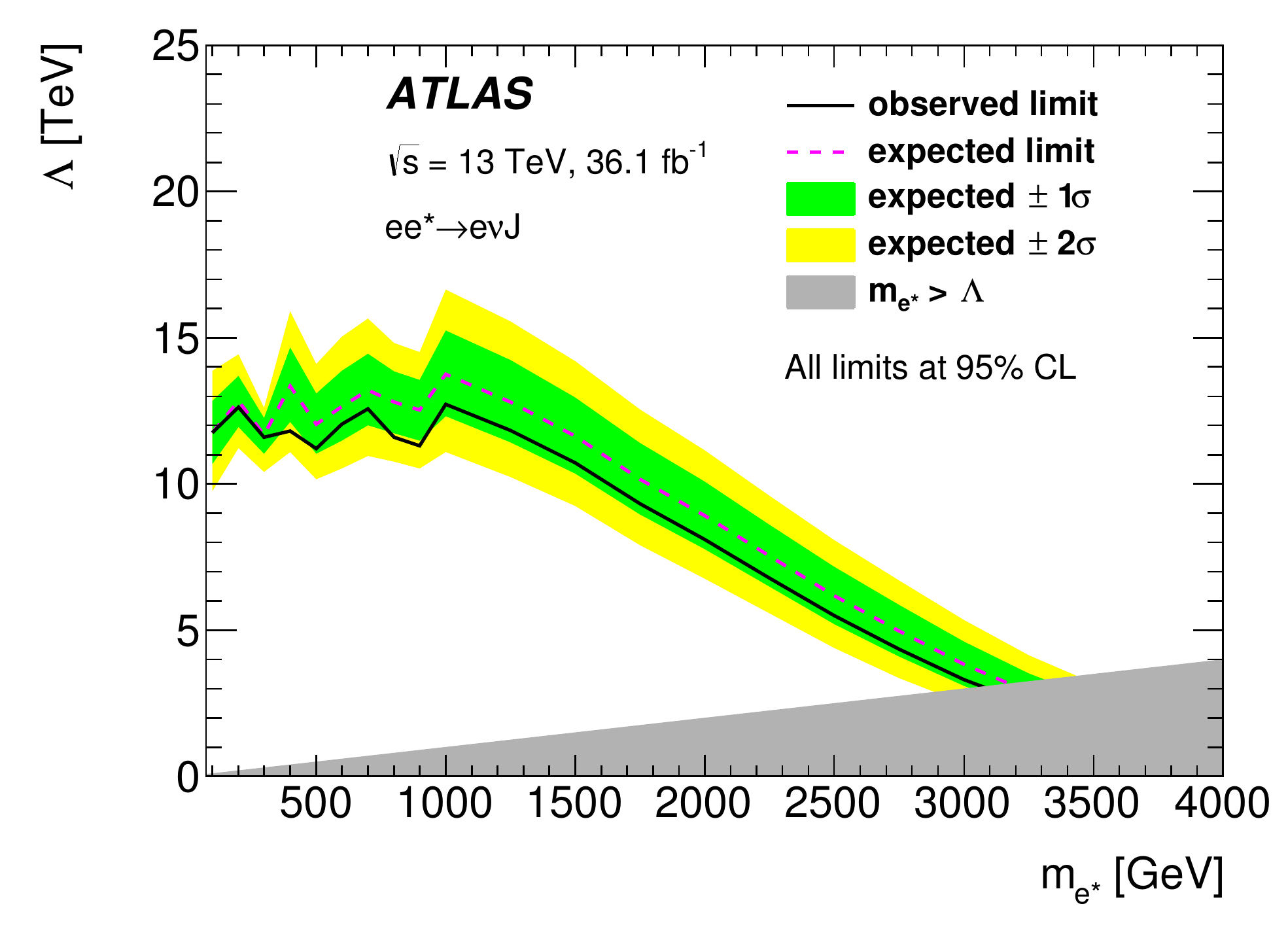}
		\caption{}
		\label{fig:enuJ:limits:la}
	\end{subfigure}
	\begin{subfigure}[t]{1\textwidth}
	\centering
		\includegraphics[width=.48\textwidth]{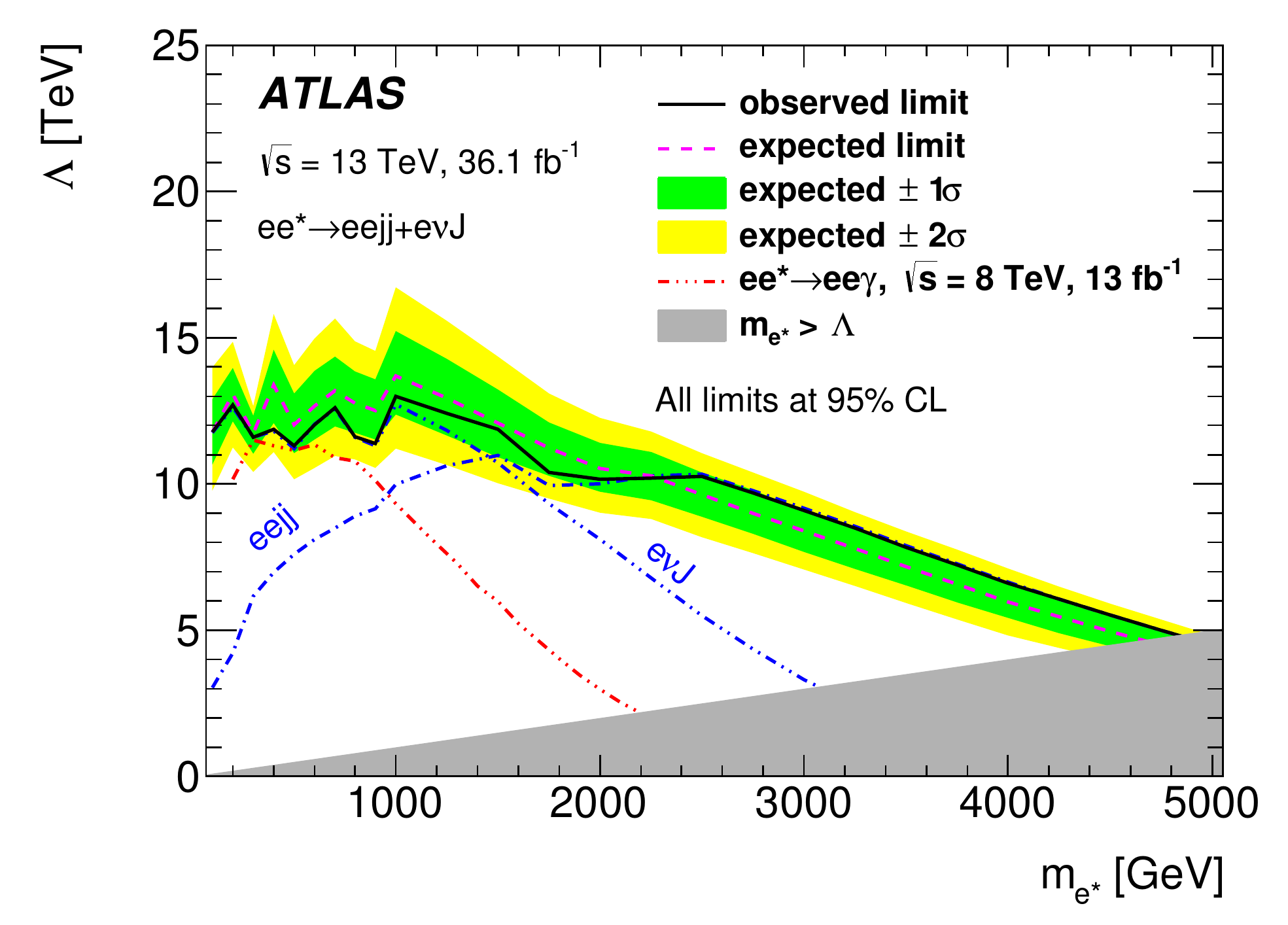}
		\caption{}
		\label{fig:comb:limits:la}
	\end{subfigure}
	\caption{Lower limits on \Lamb as a function of \mestar for (a) the \eejj channel, (b) the \enuJ channel, and (c) combined limits for both channels. 
	The $\pm1(2)\sigma$ uncertainty bands around the expected limit represent all sources of systematic and statistical uncertainties. 
	The limits for $\mestar > 4$~\TeV\ are the result of extrapolation.
	The individual observed lower limits for the \eejj (same as (a)) and the \enuJ (same as (b)) channels are shown with the blue dot-and-dash lines in  (c) for the reference.
	The exclusion limit set by ATLAS in the $ee\gamma$ search channel~\cite{Aad:2013jja} using 13~\ifb\ of data collected at \rts~=~8~\TeV\ is also shown with the red dotted line in  (c). 
	}
\end{figure}

\begin{sidewaystable}
	\caption{
Observed and expected model-independent limits at 95\% CL on the number of signal events $N_{\textrm{sig}}$ and the visible cross section $\sigma_{\textrm{vis}}$ in the signal regions of the \eejj and \enuJ channels.
The $\pm1(2)\sigma$ uncertainty intervals around the expected limit represent all sources of systematic and statistical uncertainties.
The SR10 is not defined for the \enuJ channel and denoted as ``N/A''.
	        }
	\label{tab:modindlim}
	\begin{center}
		\begin{tabular}{|l|c|c|c|c|c|c|c|c|}
			\hline
			\multirow{2}{*}{} &\multicolumn{4}{c|}{\eejj} &\multicolumn{4}{c|}{\enuJ} \\
			\cline{2-9}
			& $N^{\textrm{obs}}_{\textrm{sig}}$ & $N^{\textrm{exp}}_{\textrm{sig}}$ & $\sigma^{\textrm{obs}}_{\textrm{vis}}$ [fb] & $\sigma^{\textrm{exp}}_{\textrm{vis}}$ [fb] & $N^{\textrm{obs}}_{\textrm{sig}}$ & $N^{\textrm{exp}}_{\textrm{sig}}$ & $\sigma^{\textrm{obs}}_{\textrm{vis}}$ [fb] & $\sigma^{\textrm{exp}}_{\textrm{vis}}$ [fb]\\
			\hline 
			SR1 & $120.3$ & $107.9^{+38.1(81.6)}_{-29.1(48.8)}$ & $3.34$ & $2.99^{+1.06(2.27)}_{-0.81(1.35)}$      & $11.2$ & $11.3^{+4.6(10.5)}_{-3.1(5.2)}$ & $0.31$ & $0.31^{+0.13(0.29)}_{-0.09(0.14)}$\\
			SR2 & $30.7$ & $35.0^{+12.7(27.8)}_{-9.3(15.6)}$ & $0.85$ & $0.97^{+0.35(0.77)}_{-0.26(0.43)}$      & $21.4$ & $19.9^{+5.9(13.3)}_{-4.2(7.1)}$ & $0.59$ & $0.55^{+0.16(0.37)}_{-0.12(0.20)}$\\
			SR3 & $27.5$ & $28.7^{+11.1(24.7)}_{-8.0(13.3)}$ & $0.76$ & $0.80^{+0.31(0.69)}_{-0.22(0.37)}$      & $31.6$ & $30.9^{+7.1(15.9)}_{-5.1(6.6)}$ & $0.88$ & $0.86^{+0.20(0.45)}_{-0.14(0.21)}$\\ 
			SR4 & $18.1$ & $23.3^{+9.1(20.4)}_{-6.5(10.7)}$ & $0.50$ & $0.64^{+0.25(0.57)}_{-0.18(0.30)}$      & $23.5$ & $15.9^{+6.6(14.8)}_{-4.6(7.5)}$ & $0.65$ & $0.44^{+0.18(0.41)}_{-0.13(0.21)}$\\		
			SR5 & $13.0$ & $16.7^{+6.9(15.7)}_{-4.8(7.9)}$ & $0.36$ & $0.46^{+0.19(0.44)}_{-0.13(0.22)}$      & $26.9$ & $21.1^{+7.9(17.4)}_{-5.7(9.5)}$ & $0.75$ & $0.58^{+0.22(0.48)}_{-0.16(0.26)}$\\
			SR6 & $11.1$ & $15.8^{+6.6(15.0)}_{-4.6(7.5)}$ & $0.31$ & $0.44^{+0.18(0.42)}_{-0.13(0.21)}$      & $19.9$ & $17.0^{+6.8(15.4)}_{-4.9(8.0)}$ & $0.55$ & $0.47^{+0.19(0.43)}_{-0.13(0.22)}$\\
			SR7 & $17.0$ & $14.5^{+6.1(13.9)}_{-4.2(6.9)}$ & $0.47$ & $0.40^{+0.17(0.39)}_{-0.12(0.19)}$      & $13.8$ & $11.9^{+4.8(10.9)}_{-3.2(5.4)}$ & $0.38$ & $0.33^{+0.13(0.30)}_{-0.09(0.15)}$\\
			SR8 & $14.0$ & $14.1^{+5.9(13.4)}_{-4.0(6.7)}$ & $0.39$ & $0.39^{+0.16(0.37)}_{-0.11(0.18)}$      & $17.1$ & $11.9^{+4.6(10.5)}_{-3.1(5.2)}$ & $0.47$ & $0.33^{+0.13(0.29)}_{-0.09(0.14)}$\\
			SR9 & $9.7$ & $10.9^{+4.8(11.1)}_{-3.2(5.2)}$ & $0.27$ & $0.30^{+0.13(0.31)}_{-0.09(0.15)}$      & $9.4$ & $7.5^{+3.5(8.3)}_{-2.3(3.7)}$ & $0.26$ & $0.21^{+0.10(0.23)}_{-0.06(0.10)}$\\
			SR10 & $7.3$ & $11.1^{+4.9(11.4)}_{-3.3(5.4)}$ & $0.20$ & $0.31^{+0.14(0.32)}_{-0.09(0.15)}$      & N/A               & N/A 		& N/A                          & N/A \\
			\hline 
		\end{tabular}
	\end{center}
\end{sidewaystable}

\FloatBarrier

\section{Conclusion}\label{sec:conclusion}

A search for a singly produced excited electron in association with a SM electron is performed using \eejj and \enuJ final states with the ATLAS detector at the LHC. The search utilizes data from \pp collisions at \rts~=~13~\TeV\ with an integrated luminosity of 36.1~\ifb. No significant deviation from the SM background expectation is observed in either channel.
Upper limits are calculated for the $\pp\to\eestartoeeqqbar$ and $\pp\to\eestartoenuW$ production cross sections 
as a function of the excited electron mass \mestar at 95\% confidence level. 
Lower limits on the compositeness scale parameter \Lamb are set at 95\% confidence level as a function of \mestar. For excited electrons with $\mestar < 1.5$~\TeV, the lower limit on \Lamb is 11~\TeV, and it decreases to 7~\TeV\ at $\mestar=4$~\TeV. In the special case of the excited lepton model where $\mestar = \Lamb$, the values of $\mestar < 4.8$~\TeV\ are excluded. The sensitivity of the search is significantly better than the previous results obtained by ATLAS and CMS from LHC Run~1.
Model-independent upper limits on the number of signal events and on the visible signal cross section in the signal regions are presented. 
The latter vary between 0.20 (0.26)~fb and 3.34 (0.88)~fb for the \eejj (\enuJ) channel.

\section*{Acknowledgements}


We thank CERN for the very successful operation of the LHC, as well as the
support staff from our institutions without whom ATLAS could not be
operated efficiently.

We acknowledge the support of ANPCyT, Argentina; YerPhI, Armenia; ARC, Australia; BMWFW and FWF, Austria; ANAS, Azerbaijan; SSTC, Belarus; CNPq and FAPESP, Brazil; NSERC, NRC and CFI, Canada; CERN; CONICYT, Chile; CAS, MOST and NSFC, China; COLCIENCIAS, Colombia; MSMT CR, MPO CR and VSC CR, Czech Republic; DNRF and DNSRC, Denmark; IN2P3-CNRS, CEA-DRF/IRFU, France; SRNSFG, Georgia; BMBF, HGF, and MPG, Germany; GSRT, Greece; RGC, Hong Kong SAR, China; ISF and Benoziyo Center, Israel; INFN, Italy; MEXT and JSPS, Japan; CNRST, Morocco; NWO, Netherlands; RCN, Norway; MNiSW and NCN, Poland; FCT, Portugal; MNE/IFA, Romania; MES of Russia and NRC KI, Russian Federation; JINR; MESTD, Serbia; MSSR, Slovakia; ARRS and MIZ\v{S}, Slovenia; DST/NRF, South Africa; MINECO, Spain; SRC and Wallenberg Foundation, Sweden; SERI, SNSF and Cantons of Bern and Geneva, Switzerland; MOST, Taiwan; TAEK, Turkey; STFC, United Kingdom; DOE and NSF, United States of America. In addition, individual groups and members have received support from BCKDF, CANARIE, CRC and Compute Canada, Canada; COST, ERC, ERDF, Horizon 2020, and Marie Sk{\l}odowska-Curie Actions, European Union; Investissements d' Avenir Labex and Idex, ANR, France; DFG and AvH Foundation, Germany; Herakleitos, Thales and Aristeia programmes co-financed by EU-ESF and the Greek NSRF, Greece; BSF-NSF and GIF, Israel; CERCA Programme Generalitat de Catalunya, Spain; The Royal Society and Leverhulme Trust, United Kingdom. 

The crucial computing support from all WLCG partners is acknowledged gratefully, in particular from CERN, the ATLAS Tier-1 facilities at TRIUMF (Canada), NDGF (Denmark, Norway, Sweden), CC-IN2P3 (France), KIT/GridKA (Germany), INFN-CNAF (Italy), NL-T1 (Netherlands), PIC (Spain), ASGC (Taiwan), RAL (UK) and BNL (USA), the Tier-2 facilities worldwide and large non-WLCG resource providers. Major contributors of computing resources are listed in Ref.~\cite{ATL-GEN-PUB-2016-002}.

\clearpage

\printbibliography

\clearpage 
 
\begin{flushleft}
{\Large The ATLAS Collaboration}

\bigskip

M.~Aaboud$^\textrm{\scriptsize 35d}$,    
G.~Aad$^\textrm{\scriptsize 100}$,    
B.~Abbott$^\textrm{\scriptsize 127}$,    
D.C.~Abbott$^\textrm{\scriptsize 101}$,    
O.~Abdinov$^\textrm{\scriptsize 13,*}$,    
B.~Abeloos$^\textrm{\scriptsize 131}$,    
D.K.~Abhayasinghe$^\textrm{\scriptsize 92}$,    
S.H.~Abidi$^\textrm{\scriptsize 166}$,    
O.S.~AbouZeid$^\textrm{\scriptsize 40}$,    
N.L.~Abraham$^\textrm{\scriptsize 155}$
% ...
\bigskip
\\

% ...
$^{13}$Institute of Physics, Azerbaijan Academy of Sciences, Baku; Azerbaijan.\\
% ...
$^{35}$$^{(a)}$Facult\'e des Sciences Ain Chock, R\'eseau Universitaire de Physique des Hautes Energies - Universit\'e Hassan II, Casablanca;$^{(b)}$Facult\'{e} des Sciences, Universit\'{e} Ibn-Tofail, K\'{e}nitra;$^{(c)}$Facult\'e des Sciences Semlalia, Universit\'e Cadi Ayyad, LPHEA-Marrakech;$^{(d)}$Facult\'e des Sciences, Universit\'e Mohamed Premier and LPTPM, Oujda;$^{(e)}$Facult\'e des sciences, Universit\'e Mohammed V, Rabat; Morocco.\\
% ...
$^{40}$Niels Bohr Institute, University of Copenhagen, Copenhagen; Denmark.\\
% ...
$^{92}$Department of Physics, Royal Holloway University of London, Egham; United Kingdom.\\
% ...
$^{100}$CPPM, Aix-Marseille Universit\'e, CNRS/IN2P3, Marseille; France.\\
$^{101}$Department of Physics, University of Massachusetts, Amherst MA; United States of America.\\
% ...
$^{127}$Homer L. Dodge Department of Physics and Astronomy, University of Oklahoma, Norman OK; United States of America.\\
% ...
$^{131}$LAL, Universit\'e Paris-Sud, CNRS/IN2P3, Universit\'e Paris-Saclay, Orsay; France.\\
% ...
$^{155}$Department of Physics and Astronomy, University of Sussex, Brighton; United Kingdom.\\
% ...
$^{166}$Department of Physics, University of Toronto, Toronto ON; Canada.\\
% ...
$^{a}$ Also at Borough of Manhattan Community College, City University of New York, New York NY; United States of America.\\
$^{b}$ Also at Centre for High Performance Computing, CSIR Campus, Rosebank, Cape Town; South Africa.\\
$^{c}$ Also at CERN, Geneva; Switzerland.\\
$^{d}$ Also at CPPM, Aix-Marseille Universit\'e, CNRS/IN2P3, Marseille; France.\\
$^{e}$ Also at D\'epartement de Physique Nucl\'eaire et Corpusculaire, Universit\'e de Gen\`eve, Gen\`eve; Switzerland.\\
% ...
$^{*}$ Deceased

\end{flushleft}



\end{document}